\documentclass[12pt,preprint]{aastex}

\newcommand{\amm}{NH$_{3}$}
\newcommand{\mh}{H$_{2}$}
\newcommand{\hii}{H\,{\sc ii}}
\newcommand{\hi}{H\,{\sc i}}

\newcommand{\lum}{erg~s$^{-1}$}

\newcommand{\cden}{cm$^{-2}$}
\newcommand{\vden}{cm$^{-3}$}

\newcommand{\msun}{M$_{\sun}$}
\newcommand{\kms}{km~s$^{-1}$}

\slugcomment{}

\shorttitle{Ammonia in NGC\,253}
\shortauthors{Ott et al.}

\begin{document}

\title{The Temperature Distribution of Dense Molecular Gas in the Center of NGC\,253}


\author{J{\"u}rgen Ott\footnote{Bolton Fellow}}
\affil{CSIRO Australia Telescope National Facility, Cnr Vimiera \& Pembroke Roads, Marsfield NSW 2122, Australia}
\email{Juergen.Ott@csiro.au}

\author{Axel Wei{\ss}}
\affil{Instituto de Radio Astronom{\'{\i}}a Milim{\'e}trica, Avenida Divina Pastora 7, 18012 Granada, Spain}
\email{aweiss@iram.es}

\author{Christian Henkel}
\affil{Max-Planck-Institut f{\"u}r Radioastronomie, Auf dem H{\"u}gel 69, 53121 Bonn, Germany}
\email{chenkel@mpifr-bonn.mpg.de}

\and

\author{Fabian Walter}
\affil{Max-Planck-Institut f{\"u}r Astronomie, K{\"o}nigstuhl 17, 69117 Heidelberg, Germany}
\email{walter@mpia.de}

\begin{abstract}

We present interferometric maps of ammonia (\amm) of the nearby
starburst galaxy NGC\,253. The observations have been taken with the
Australia Telescope Compact Array and include the para-\amm\ (1,1),
(2,2), and the ortho-\amm\ (3,3) and (6,6) inversion lines. Six major
complexes of dense ammonia are identified, three of them on either
side of the starburst center, out to projected galactocentric radii of
$\sim 250$\,pc. Rotational temperatures are derived toward selected
individual positions as well as for the entire south-eastern and
north-western molecular complexes. The application of radiative
transfer large velocity gradient models reveals that the bulk of the
ammonia molecules is embedded in a one-temperature gas phase. Kinetic
temperatures of this gas are $\sim$200 and 140\,K toward the
south-west and north-east, respectively. The temperatures under which
ammonia was formed in the past are with $\gtrsim$30\,K also warmer
toward the south-west than toward the north-east ($\sim
15-20$\,K). This is indicated by the ortho--to--para ammonia ratio
which is $\sim 1$ and $1.5-2.5$ toward the south-west and north-east,
respectively. Ammonia column densities in the brightest complexes are
in the range of $6-11\times10^{14}$\,\cden, which adds up to a total
ammonia mass of $\sim 20$\,M$_{\sun}$, about evenly distributed toward
both sides of the nucleus. Ammonia abundances relative to \mh\ are
$\sim 3\times 10^{-8}$. In the south--western complex, the ammonia
abundances increase from the starburst center to larger galactocentric
radii. Toward the center of NGC\,253, \amm (1,1), (2,2), and (6,6) is
detected in absorption against an unresolved continuum source. At the
same location, however, ammonia (3,3) is found in emission which
indicates maser activity. This would be the first detected
extragalactic \amm\ maser. Evidence for an expanding shell in the
south-western complex is provided. The shell, with a dynamical age of
$\sim 1.3$\,Myr, is centered on an X-ray point source which must be
located within the dense gas of NGC\,253. The shell and X-ray
properties can be reproduced by the energy input of a highly obscured
young stellar cluster with a mass of $\sim 10^{5}$\,M$_{\sun}$ which
also heats the dense gas. A current star formation rate of $\sim
2.8$\,M$_{\sun}$\,yr$^{-1}$ is derived for the nuclear starburst in
NGC\,253 based on its 1.2\,cm continuum emission.

\end{abstract}

\object{NGC 253}

\section{Introduction}
\label{sec:intro}

Starburst galaxies are a main contributor to the star formation (SF)
in the universe \citep[$\sim 20$\% at $z<0.2$;][]{bri04}. The fuel for
star formation is dense molecular gas. Recently \citet{gao04} showed
that in actively star forming galaxies the molecular gas with volume
densities $> 10^{4}$\,\vden\ (traced by HCN) correlates very well with
the SF rate (SFR) determined by the far-infrared (FIR) emission. Less
dense molecular gas, such as that traced by CO, is a weaker measure of
the current SFR. The physical state of the molecular gas is mainly
described by its temperature and density. It is thus important to
constrain the range of those two fundamental parameters to understand
how molecular gas feeds regions of massive SF in nuclear environments.

Tracers for high density molecular gas are transitions of molecules
with large electric dipole moments (relative to CO), such as HCN,
HCO$^{+}$, CS, or ammonia (\amm). Those molecules are excited at
densities $\gtrsim 10^{4}$\,\vden, which we refer to as {\it dense}
gas. The specific tetrahedral structure of ammonia that permits the
tunneling of the nitrogen atom through the plane defined by the three
hydrogen atoms causes inversion doublets. Since the energy difference
between the two states of a given inversion doublet does only weakly
depend on the rotational quantum numbers $J$ and $K$, a vast range of
molecular excitation can be covered with just one set of radio
receivers.

For the metastable ($J$=$K$) inversion lines, at the bottom of each
$K$-ladder, radiative decay is extremely slow and the rotation
temperature $T_{\rm rot}$ (the temperature defined by the populations
of the different rotational levels) is to first order similar to the
kinetic temperature ($T_{\rm kin}$) of the dense molecular gas
\citep[e.g.,][]{ho83,wal83,stu85,mar86,ung86,dan88,flo95,hue95}.
Deviations between $T_{\rm rot}$ and $T_{\rm kin}$ at higher
temperatures are well described by radiative transfer large velocity
gradient (LVG) models. Ammonia is the brightest molecular species with
those properties which makes it an important and easy-to-use
thermometer. According to the orientation of the hydrogen spins, two
different variants of ammonia can be distinguished: ortho-\amm\ (all
three hydrogen spins are parallel) and para-\amm\ (one spin is
anti-parallel). This restricts the rotational ($J,K$) states to $K=3n$
($n$ is an integer, $J$ the quantum number of the total angular
momentum, $K$ its projection onto the symmetry axis of the molecule)
for ortho and $K\neq3n$ for para-ammonia
\citep[e.g.,][]{tow55,ho83}. Since the lowest state of ammonia belongs
to ortho-ammonia, an ortho--to--para-ammonia ratio exceeding unity is
expected if the formation and equilibration of \amm\ is performed in
an environment with a low energy content \citep[e.g., ][]{tak02}.

Up to now, rotational ammonia temperatures in extragalactic objects
were only derived from single dish observations
\citep[e.g., ][]{mar86,hen00,tak00,wei01a,tak02,mau03} which restricted the
temperature measurements to global averages over a few hundred pc. The
only interferometric (VLA) extragalactic ammonia maps were presented
by \citet{ho90} for IC\,342. The signal-to-noise ratio of these maps,
however, was not good enough for temperature determinations. In this
paper we present interferometric maps of ammonia toward the core of
the galaxy NGC\,253 observed with the Australia Telescope Compact
Array (ATCA). The observations have a linear resolution of $\sim 5
\arcsec$ (corresponding to 65\,pc) along the major axis and are
therefore a clear improvement over single dish observations presented
by \citet{tak02} (hereafter T02) and \citet{mau03} (M03). NGC\,253 is
one of the most nearby starburst galaxies (distance adopted here:
2.6\,Mpc, \citealt{puc88}; but see \citealt{kar03}: 3.9\,Mpc). The
current SFR of $\sim 3$\,\msun\,yr$^{-1}$ \citep{rad01} is mainly
concentrated in the inner $\sim 200$\,pc. The starburst, likely fed by
gas streaming toward the center along a prominent bar, is visible at
virtually all wavelengths
\citep[e.g.,][]{ulv97,eng98,for00,str02,jar03}. 
Furthermore, feedback from massive stars in the form of strong stellar
winds and supernova explosions heats the surrounding gas and a
galactic wind is observable in NGC\,253 up to 8\,kpc above the disk
\citep{str02}. Molecular lines have been abundantly found toward the
nucleus of NGC\,253
\citep[e.g.,][]{tur85,jac95,hou97,gar00,bra03,mau03,mar03,pag04,hen04,bay04} 
and, in fact, almost every molecule which was detected in an
extragalactic object has also been detected in NGC\,253.

In Sect.\,\ref{sec:obs} we describe the observational parameters of
our ATCA ammonia data as well as the data reduction techniques. This
is followed by the presentation of the data in
Sect.\,\ref{sec:results}. We discuss the ammonia and continuum
observations in Sect.\,\ref{sec:discuss} and summarize the paper in
Sect.\,\ref{sec:summary}.

\section{Observations and Data Analysis}
\label{sec:obs}

Observations of the para-\amm\ (1,1) and (2,2) inversion lines were
simultaneously performed with the ATCA on 2003 September 05. On 2004
April 26, the ATCA was used to simultaneously observe ortho-\amm\
(3,3) and (6,6). During both observations, the ATCA was in the compact
EW\,367 array configuration. The bandwidths were chosen to be 64\,MHz
for each line, centered on the redshifted frequencies of 23.6755,
23.7035, 23.851, and 25.036\,GHz, respectively (primary half-power
beam widths: $\sim 2\farcm2$). Each of the frequency bands were split
into 64 channels resulting in a channel width of 1\,MHz or $\sim
12.65$\,\kms\ ($\sim 11.96$\,\kms\ in the case of the \amm\ (6,6)
line). For the calibration of the \amm\ (1,1) and (2,2) observations
we used Uranus as a flux calibrator, PKS\,1921-293 as a bandpass
calibrator (integration time $\sim 10$\,min), and PKS\,0023-263 as a
phase calibrator (4\,min after each 20\,min observing interval on
NGC\,253). The calibrators for the \amm\ (3,3) and (6,6) observations
were the same except that PKS\,1934-638 was used to determine the flux
scales. Calibration uncertainties are estimated to be $\sim 10$\%. Due
to the lower signal--to--noise ratio of the \amm (6,6) line, we
estimate the error for this line to $\sim 20$\%. Relative
uncertainties between the different line fluxes do not depend on
absolute calibration errors and are $\sim 5$\%. The total \amm\ (1,1)
and (2,2) integration time on NGC\,253 was $\sim 8.5$\,h spread over a
12\,h interval yielding a good uv-coverage of the east-west
interferometer. The \amm\ (3,3) and (6,6) integration time was $\sim
6$ hours spread over $\sim 7$ hours. This resulted in a reduced
uv-coverage. Both, the XX and the YY polarizations were observed at
all frequencies.

The data were reduced with the {\sc MIRIAD} \citep{sau95} and
visualized with the KARMA \citep{goo96} software packages. After
flagging very few corrupted visibilities and discarding edge channels,
calibration was applied to the data. Subsequently, the data were
Fourier transformed to image data cubes with a gridding of 1\arcsec\
pixel size. Along with the transform, we used natural weighting which
results in a synthesized beam of $18\farcs8\times5\farcs2$ in size
(position angle: $-0.7\degr$) for the full uv-coverage \amm\
(1,1)/(2,2) data. Continuum maps were produced by averaging line-free
channels. Due to sidelobes of the strong emission, the rms noise of
the naturally weighted continuum maps is with $\sim
1$\,mJy\,beam$^{-1}$ at 23.6755 and 23.7035\,GHz and $\sim
1.5$\,mJy\,beam$^{-1}$ at 23.851 and 25.036\,GHz about twice that of
the continuum-subtracted channel maps (see below). The continuum maps
were subtracted from all channel maps to obtain continuum-free line
cubes. Since the uv-coverages of the \amm\ (1,1)/(2,2) and \amm\
(3,3)/(6,6) cubes differ we also produced (1,1) and (2,2) data cubes
restricted to the hour angles set by the \amm\ (3,3)/(6,6)
observations. This resulted in a synthesized beam of
$30\farcs41\times5\farcs09$ and a position angle of $-12\degr$, which
is almost orthogonal to the distribution of the dense gas as seen,
e.g., in \citet{pen96}. The data was Hanning smoothed (which resulted
in a velocity resolution about twice the channel width) and {\sc
CLEAN}ed down to a 1\,mJy level (about $2\sigma$ rms noise).

For a proper comparison of the line properties, the \amm\ (6,6) and
the LST restricted (1,1)/(2,2) cubes were finally smoothed to the
spatial resolution of the \amm\ (3,3) data. The channel maps of all
data cubes exhibit an rms noise of $\sim 0.6$\,mJy\,beam$^{-1}$. We
also produced a super-resolved cube of \amm\ (3,3) by convolving the
{\sc CLEAN} components with a circular beam of
$5\arcsec\times5\arcsec$. The super-resolved cube is not used for any
quantitative analysis but provides a better morphological separation
of individual clumps of dense molecular gas. Moment maps were computed
by selecting only emission components with a signal at least 2.5 times
above the rms per channel map. In addition, the signal must be visible
in at least three consecutive channels to be incorporated. All other
data was flagged before producing the integrated intensity, the
intensity weighted velocity, and the velocity dispersion maps (i.e.,
Moments 0, 1, and 2).

\section{Results}
\label{sec:results}

\subsection{Continuum Emission}
\label{sec:cont}

Spatially unresolved continuum emission was detected in all bands
toward the central starburst of NGC\,253 at
$\alpha_{(J2000)}=00^{h}\,47^{m}\,33\fs2$,
$\delta_{(J2000)}=-25\degr\,17\arcmin\,17\arcsec$. The flux density of the
continuum source is derived to $520$\,mJy with an absolute uncertainty
of $\sim 10$\%. The flux is in good agreement with the 31.4\,GHz flux
of $(590\pm80)$\,mJy reported by
\citet{gel81}. The relative uncertainty of the continuum fluxes in the
four different bands (assuming a flat spectrum within the narrow
frequency range) is only $\sim 5$\%.

\subsection{Global Morphology and Velocity Field of the Ammonia Inversion Lines}
\label{sec:morph}

\begin{figure}
\epsscale{1}

\plotone{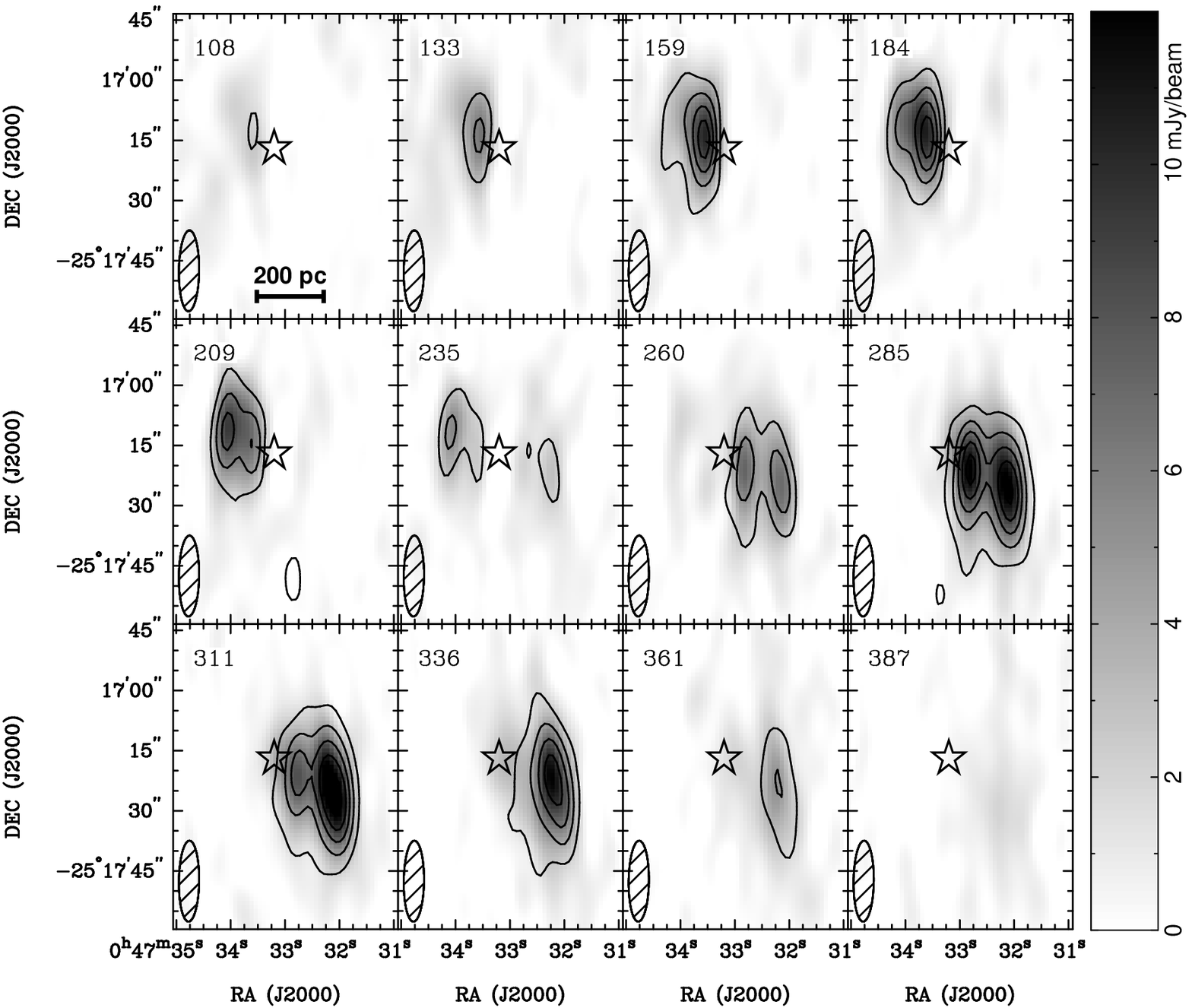}

\caption
{Channel maps of the naturally weighted \amm (1,1) observations of the
central region of NGC\,253 with the \amm (2,2) overlaid as contours
(greyscale ranging from 0--12\,mJy\,beam$^{-1}$; contours start at and
are spaced by 2\,mJy\,beam$^{-1}$). The star marks the peak of the
1.2\,cm continuum which is coincident with the starburst center. The
synthesized beam of the observations is shown in the lower left corner
of each panel and the velocity is given in units of
\kms\ in the upper left corners. Each plot is an average over two
channels which is about the velocity resolution of the Hanning smoothed data.
\label{fig:chanmaps11}}
\end{figure}

\placefigure{fig:chanmaps11}

\begin{figure}
\epsscale{1}

\plotone{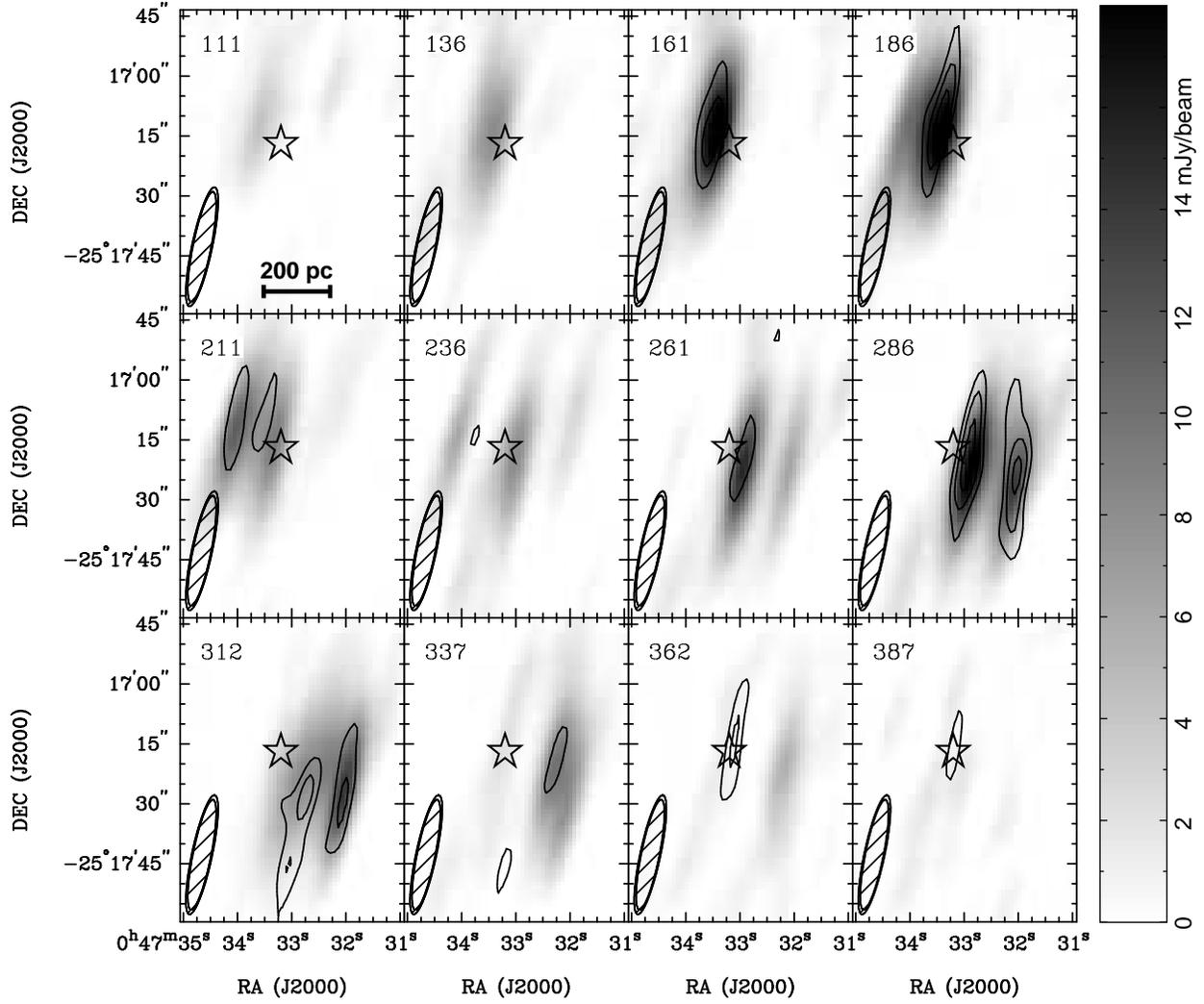}

\caption
{Channel maps of the naturally weighted \amm (3,3) emission with
\amm (6,6) overlaid as contours (see also Fig.\,\ref{fig:chanmaps11} 
for the symbols). Here the greyscale ranges from 0 to
18\,mJy\,beam$^{-1}$ and the contours start at 2\,mJy\,beam$^{-1}$ and
are spaced by 1\,mJy\,beam$^{-1}$.
\label{fig:chanmaps33}}
\end{figure}

\placefigure{fig:chanmaps33}

\begin{figure}
\epsscale{1}

\plotone{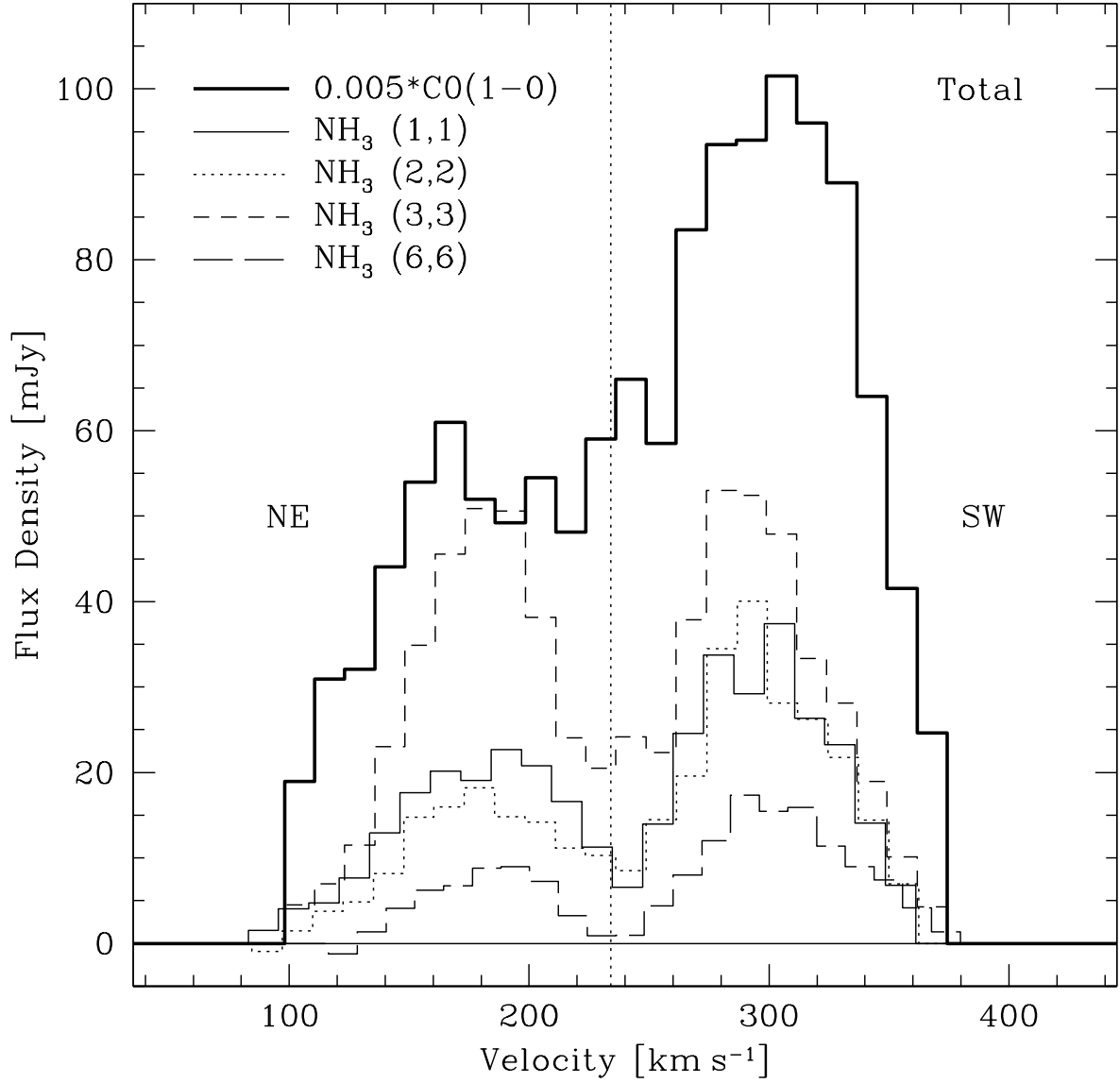}
\caption
{Total spectra of the ammonia inversion lines (see
Fig.\,\ref{fig:moments} for the corresponding maps) and the scaled
(scaling factor 0.005) CO(1-0) transition in the inner $\sim 600$\,pc
of NGC\,253. The {\it vertical line} marks the systemic velocity of
NGC\,253.
\label{fig:totspec}}
\end{figure}

\begin{figure}
\epsscale{0.8}

\plotone{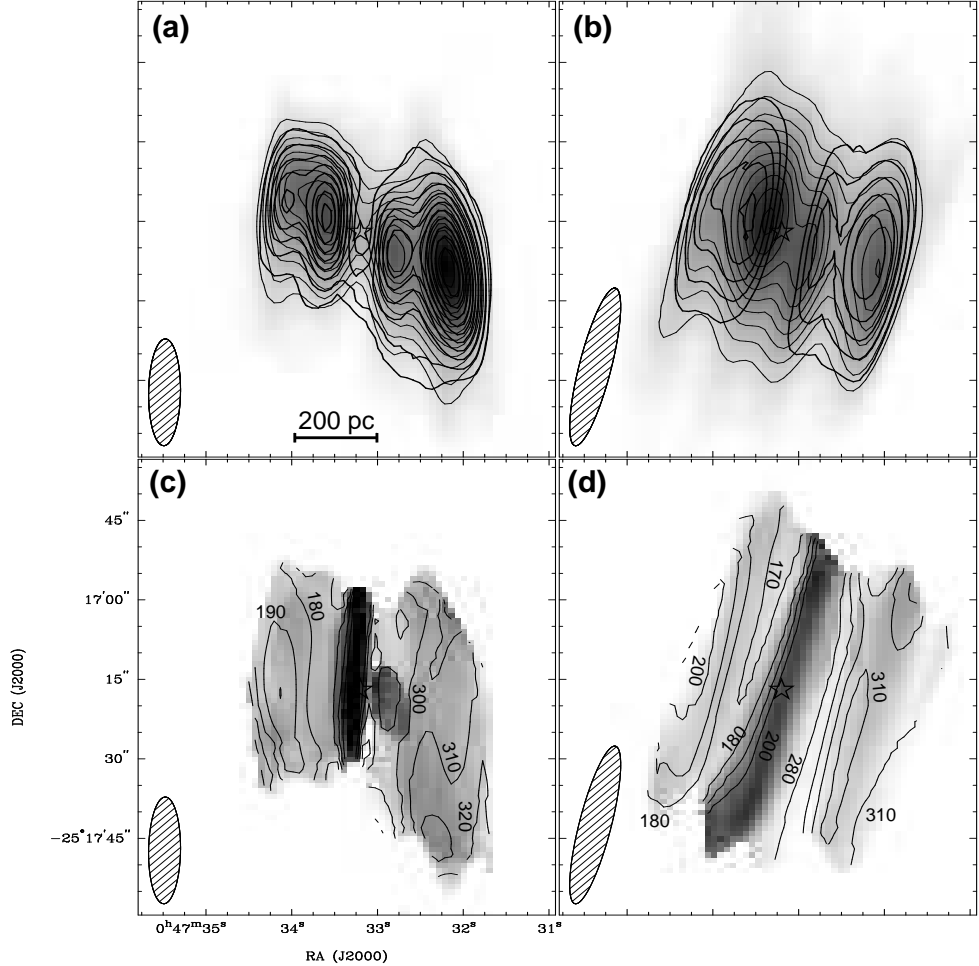}
\caption
{Moment maps of the ammonia emission in NGC\,253. {\bf (a)} Naturally
weighted integrated intensity \amm (1,1) emission, overlaid with its
own contours ({\it thin}) and with contours of the \amm (2,2)
emission ({\it thick}). The data have been reduced using the full 12
hour uv-coverage. The greyscale is from 0 to 30\,K\,\kms\ and the
contours start at 4\,K\,\kms\ and are spaced by 2\,K\,\kms. {\bf (b)}
The \amm (3,3) integrated emission again overlaid with its own
contours ({\it thin}) and with contours of the restricted uv-coverage
\amm (1,1) emission ({\it thick} contours) in order to achieve the
same beam size and position angle. The contours are at the same
spacing and scale as in (a) and the greyscale ranges from 0 to
25\,K\,\kms. {\bf (c)} Velocity dispersion map of the \amm (1,1)
emission (full 12\,h coverage) overlaid with contours of its velocity
field. The greyscale range is $20-70$\,\kms\ and the contours of the
velocity field are spaced by 10\,\kms. Note that, for clarity, we do
not show velocity contours in the central $210-270$\,\kms\ range. {\bf
(d)} The same as (c) but for the \amm (3,3) emission. In all panels,
the position of the 1.2\,cm continuum source is marked by a {\it star}
and the corresponding beams are shown in the lower left corners. All
panels are on the same scale.
\label{fig:moments}}
\end{figure}

The resulting ammonia data cubes are shown as channel maps in
Figs.\,\ref{fig:chanmaps11} and \ref{fig:chanmaps33}. Note that for
these figures the full uv-coverage data of the \amm\ (1,1) and (2,2)
data are used. All ammonia lines are detected and two major complexes,
separated spatially and in velocity space can be defined: one toward
the north-east (NE) and a second one toward the south-west (SW) of the
starburst center. At the starburst center itself, a faint absorption
component in \amm\ (1,1), (2,2), and (6,6) and some broad (3,3)
emission are detected against the 1.2\,cm radio continuum
(Sect.\,\ref{sec:cont}). The NE component covers a velocity of $\sim
100-240$\,\kms\ in the LSR system \citep[systemic LSR velocity of
NGC\,253: $234\pm2$\,\kms;][]{whi99} whereas the SW part covers $\sim
240-380$\,\kms. The total \amm\ spectra, encompassing both the SW and
the NE components, are shown in Fig.\,\ref{fig:totspec} and their
properties are listed in Table\,\ref{tab:flux}. A comparison with the
single dish observations of M03 and T02 shows that all flux has been
recovered by the ATCA observations. In Fig.\,\ref{fig:totspec} we
overlay a CO spectrum for comparison. The CO data was observed with
the Owens Valley Radio Interferometer (courtesy of M. Dahlem and
F. Walter) and was smoothed to the ATCA beam of our observations. The
\amm\ spectra of NGC\,253 are very similar to those of CO with the
high velocity SW component to be about twice as bright as the lower
velocity NE component. However, the spectra deviate near the systemic
velocity, i.e., close to the starburst center where ammonia appears to
be less abundant than at larger galactocentric radii (see
Sect.\,\ref{sec:n}). In addition, the \amm\ (3,3) line exhibits two
components with similar flux levels. This behavior is also visible in
the integrated intensity maps shown in Fig.\,\ref{fig:moments}. In
addition to the spectral similarities, the morphologies of the \amm\
(1,1), (2,2), and (6,6) inversion lines are very similar, too. The
most central NE \amm\ (3,3) peak, however, is shifted slightly closer
to the starburst center as compared to the peaks seen in all other
inversion lines. The separation of the (3,3) emission from the other
inversion lines in that region is about the extent of the minor axis
beam ($\sim 5\arcsec$) and is therefore significant
(Fig.\,\ref{fig:moments}[b]). Such spectral and morphological
differences of the \amm\ (3,3) line are somewhat surprising, as both
the (3,3) and (6,6) lines are emitted by ortho-\amm\ (see
Sect.\,\ref{sec:clumpprop}).

\placefigure{fig:totspec}

\placetable{tab:flux}

Toward the SW region of NGC\,253 the intensity weighted, first moment
velocity field is rather regular and increases with rising
galactocentric distance from $v_{\rm LSR}\sim 240$ to $380$\,\kms. The
situation is somewhat different in the NE; the \amm\ velocities
decline from $v\sim 200$ to $150$\,\kms\ at a distance of about
8\arcsec\ from the central starburst region and then rise again to
values of $\sim 200$\,\kms. This trend is already visible in the
intensity weighted velocity maps in Fig.\,\ref{fig:moments} but
becomes more prominent in the position-velocity (pV) diagram shown in
Fig.\,\ref{fig:pv}. Linewidths are $\sim 40-90$\,\kms, irrespective of
galactocentric radius (see Fig.\,\ref{fig:moments} and
Table\,\ref{tab:clumps}).

\placefigure{fig:moments}

\section{Discussion}
\label{sec:discuss}

\subsection{The Star Formation Rate in the Nucleus of NGC\,253}

\citet{ulv97} showed that most of the radio
continuum in the center of NGC\,253 is emitted by \hii\ regions and
supernova remnants and that no clear evidence for an AGN
exists. Therefore, the measured 1.2\,cm radio continuum emission can
be used to estimate SFRs. To do so, we use the equations given in
\citet{haa00} \citep[see also][]{con92} and, using the 1.2\,cm
continuum flux of $(520\pm52)$\,mJy, we derive the nuclear SFR of
NGC\,253 to $(2.8\pm0.3)$\,\msun\,yr$^{-1}$. This is in excellent
agreement with far-infrared data (SFR$_{\rm
FIR}$=2.6--3.3\,\msun\,yr$^{-1}$;
\citealt{rad01}).

\subsection{Properties of Individual Dense Gas Clumps}
\label{sec:clumpprop}

\begin{figure}
\epsscale{0.55}

\plotone{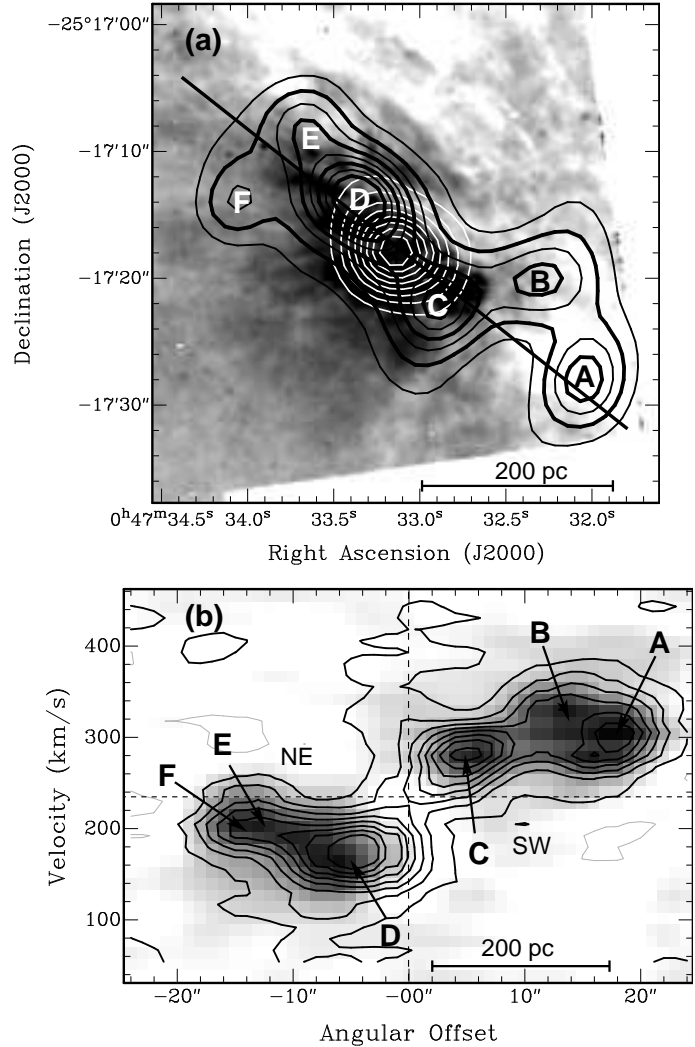}
\caption
{{\bf (a)} Super-resolved \amm (3,3) emission (resolution:
$5\arcsec\times5\arcsec$) displayed as {\it black contours} on a
logarithmic HST/WFPC2 F814W image of the core of NGC\,253. The contour
levels start at and are spaced by 10\,K\,\kms. The {\it white}
contours show the 1.2\,cm continuum emission. A position--velocity
diagram along the major axis ({\it black diagonal line}) of the
restricted uv-coverage \amm (1,1) data with contours of \amm (3,3)
(contours start from 4\% of the peak flux in steps of 10\%) is shown
in panel {\bf (b)}. Note that the data in the pV diagram are {\it not}
super-resolved and, due to the elongated beam (see
Fig.\,\ref{fig:moments}), all the molecular complexes are visible in
this diagram even if they are offset from the black diagonal line in
(a). The systemic velocity of NGC\,253 and the position of the
starburst center are shown as {\it dashed horizontal and vertical
lines}, respectively. Positive angular offsets are toward the SW.
\label{fig:pv}}
\end{figure}

In the super-resolved data cubes (Fig.\,\ref{fig:pv}[a]) six clumps
can be identified: clumps A, B, and C are located toward the SW and
clumps D, E, and F toward the NE. Only clumps C and D, however, are
symmetrical in position and velocity to the starburst center. No
symmetry is observed between complexes E/F and A/B. This indicates
that an inner ring rotating like a solid body may exist but no such
structure can explain the velocity field at larger radii of $\sim
200$\,pc. The SiO and H$^{13}$CO$^{+}$ distributions are similar to
that of \amm\ and \citet{gar00} interpret the two inner clumps being
at the inner Lindblad resonance whereas the outer clumps are the
vertices of the trailing spiral arms across the outer Lindblad
resonance. This interpretation can potentially explain the kinematic
asymmetry of the outer parts of the dense gas. The morphology and
kinematics of the \amm\ complexes are also in general agreement with
CS data (a different tracer for dense gas) presented in
\citet{pen96}. Clumps with CS emission which may be attributed to the 
ammonia complexes E and F, however, are shifted $\sim 5\arcsec$ toward
the center with respect to \amm. \citet{pen96} show that some CS
emission in the SW, the receding part of NGC\,253, is below the
systemic velocity. Toward the approaching NE region some gas is found
with velocities above the systemic velocity. They explain those
features as signatures ($x_{1}$ and $x_{2}$ orbits) of dense gas
moving in a bar potential \citep[see also][]{sor00,das01}. Our
\amm\ data do not show those specific signatures, which, however, may
be due to the weakness of the emission toward those regions.

\placetable{tab:clumps}

\placefigure{fig:pv}

\begin{figure}
\epsscale{1}

\plotone{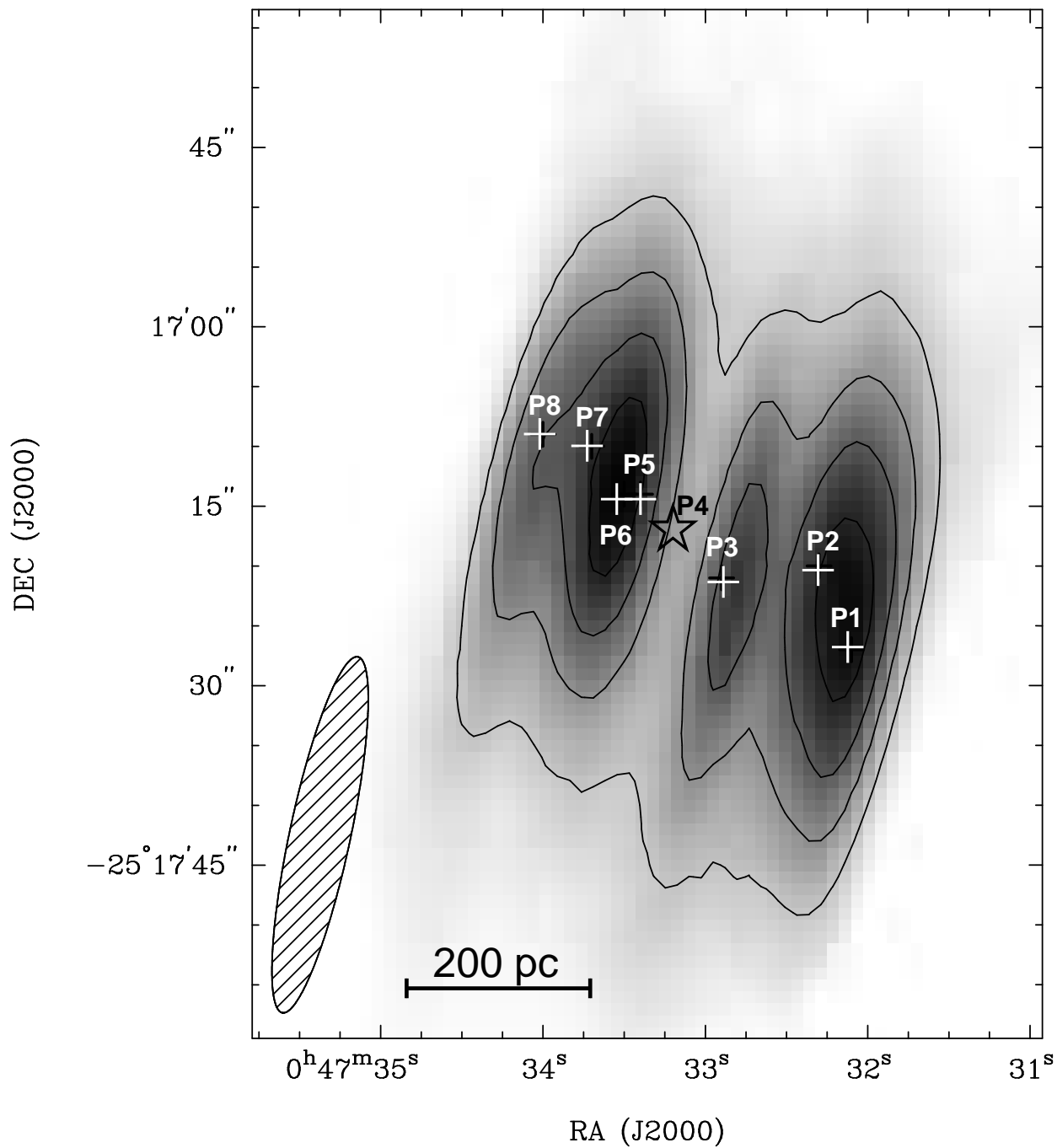}
\caption
{Total ammonia column density map and contours, computed by the sum of
the column densities derived from \amm (1,1), (2,2), (3,3), and (6,6)
[$N_{\rm 1236}({\rm NH_{3}})$]. The contour levels start at and are
spaced by $1\times10^{14}$\,\cden. Overlaid are the locations of the
positions P1 to P8 as {\it white crosses} and the 1.2\,cm continuum
peak, which corresponds to P4 as a {\it black star}. The size of the
beam is shown in the lower left corner.
\label{fig:ntot}}

\end{figure}

\begin{figure}
\epsscale{1}

\plotone{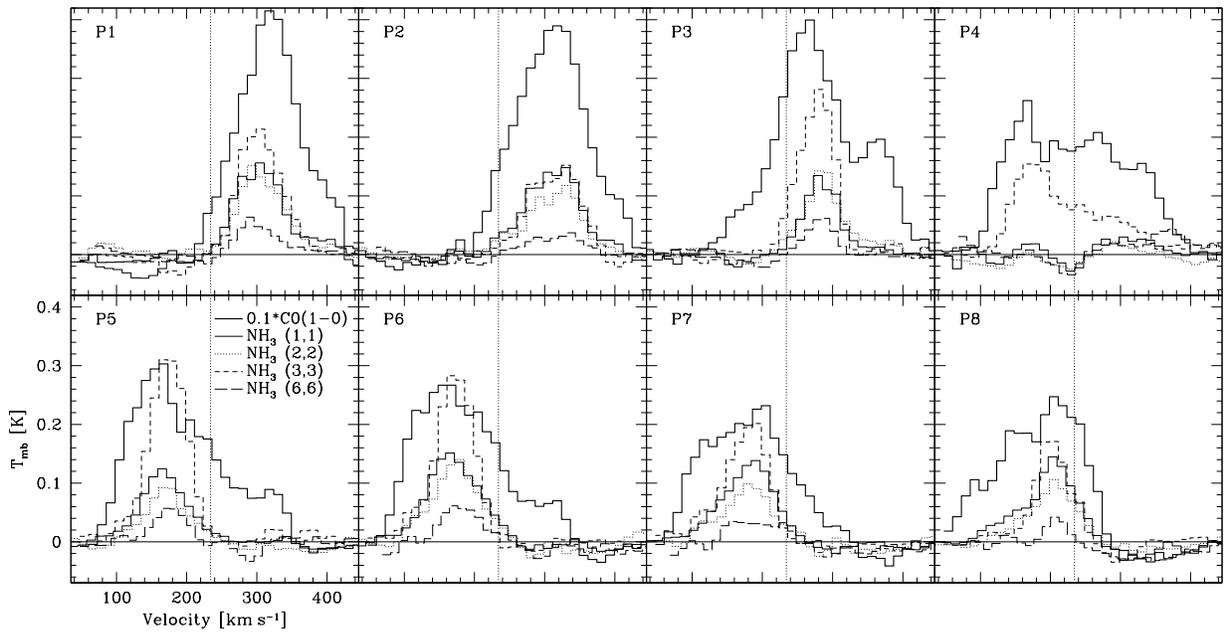}
\caption
{Ammonia spectra toward positions P1 to P8. Overlaid are corresponding
CO spectra scaled by a factor of 0.1 (CO data courtesy of M. Dahlem
and F. Walter ). Dotted {\it vertical lines} mark the systemic
velocity of NGC\,253.
\label{fig:pos}}

\end{figure}

Measured from the starburst center, the ammonia emission extends $\sim
200$\,pc toward the NE and $\sim 250$\,pc toward the SW. We define
eight positions for a more detailed analysis: Positions P1 to P3
correspond to the centers of clumps A, B, and C and P4 is located on
the 1.2\,cm continuum emission. In order to account for the small
displacement of the \amm\ (3,3) emission with respect to the other
inversion lines in the innermost NE complex, P5 is defined to be
toward the \amm\ (3,3) peak and P6 coincides with the peaks of \amm
(1,1), (2,2), and (6,6). Finally, P7 and P8 are defined to be on the
respective peaks of clumps E and F. The positions of P1 to P8 are
displayed in Fig.\,\ref{fig:ntot} and their line properties are listed
in Table\,\ref{tab:clumps}. Ammonia spectra of those positions are
shown in Fig.\,\ref{fig:pos}. In the same figure, we also show CO
spectra for comparison. The individual ammonia linewidths and
brightness peaks toward the SW do not differ significantly from those
toward the NE region despite the larger total, spatially integrated
flux of the SW region (see Sect.\,\ref{sec:morph},
Fig.\,\ref{fig:totspec}, and Table\,\ref{tab:flux}). This implies that
there is more extended emission toward the SW than the NE. The CO
spectra for P4 to P7 are very broad and some subcomponents extend well
to the other side of the systemic velocity of NGC\,253. This is
similar to the velocity structure of CS \citep{pen96} but, as noted
above, cannot be extracted from our ammonia data.

\amm\ (1,1), (2,2), and (6,6) absorption is seen toward the 1.2\,cm 
continuum source at P4. If a similar absorption feature for \amm\
(3,3) exists, it is blended with emission of this line. The reason for
the \amm\ (3,3) emission at P4 may be obvious from the contours of the
pV diagram in Fig.\,\ref{fig:pv}, where a faint, central \amm\ (3,3)
feature extends from the systemic velocity of NGC\,253 to $\sim
400$\,\kms. A similar feature is seen in OH \citep{tur85} and SiO
\citep{gar00}. \citet{gar00} speculate that this may be evidence for 
molecular outflows from the starburst. Our elongated beam toward the
possible outflow direction, however, inhibits the identification of a
spatial offset of the faint plume along the minor axis of
NGC\,253. Therefore, we cannot exclude other line broadening
mechanisms. The reason why we only observe this feature in \amm\ (3,3)
but not in the other ammonia transitions may be the peculiar
distribution of energy levels in ortho-\amm, which allow the line to
be a maser under certain circumstances. \citet{wal83} show that at
\mh\ volume densities of $10^{4-5}$\,\vden\ the populations of the two
inversion levels can be inverted \citep[see also][]{sch89}. Such
masers are known to exist in a few prominent Galactic star forming
regions, such as DR\,21 and W\,33
\citep[][]{gui83,wil90,wil95}. Apparently, these masers are
unsaturated, requiring seed photons from the background continuum of
the associated \hii\ regions. It may therefore not be accidental that
the \amm\ (3,3) emission feature, accompanied by (1,1), (2,2), and
(6,6) absorption features, is observed toward the nuclear continuum of
NGC\,253. Toward the same central region of NGC\,253 OH and H$_{2}$O
masers are already known to exist \citep[][]{fra98,hen04}. Thus, \amm\
may represent the third molecular species known to exhibit maser
emission in the central environment of NGC\,253. If confirmed, it
would be the first extragalactic \amm\ maser ever detected
\citep[\amm\ would then be the sixth molecule showing maser emission
in extragalactic objects; for OH, H$_{2}$O, CH, H$_{2}$CO, and SiO,
see][]{wel71,chu77,whi80,baa86,vlo96}.

\placefigure{fig:ntot}

\placefigure{fig:pos}

\subsection{Rotational Temperatures}
\label{sec:Trot}

As indicated in the introduction, ammonia can be used to determine
rotational temperatures of the dense molecular gas. To derive
rotational temperatures from ammonia in emission, populations of more
than one metastable ($J=K$) inversion doublet have to be
determined. Assuming optically thin line emission, the column
densities of their upper states $N_{\rm u}$ can be derived via:

\begin{equation}
N_{u}(J,K)=\frac{7.77\times10^{13}}{\nu}\frac{J(J+1)}{K^{2}}\,\,\,\int
T_{\rm mb}\,\, dv
\label{eq:n}
\end{equation}

\noindent \citep[][]{hen00} where $N_{\rm u}$ is given in units of \cden, the 
frequency $\nu$ in GHz, the main beam brightness temperature $T_{\rm
mb}$ in K, and the velocity $v$ in \kms.

For ammonia in absorption, the column densities of the different
populations cannot be determined without the knowledge of the
excitation temperature $T_{\rm ex}$ across an individual inversion
doublet. The following equation applies:

\begin{equation}
\frac{N(J,K)}{T_{ex}}=1.61\times 10^{14}\,\, \frac{J(J+1)}{K^{2}\nu}\,\, \tau\,\Delta v_{1/2}
\label{eq:abs}
\end{equation}

\noindent \citep{hue95} with $\Delta v_{1/2}$ denoting the FWHM of the line 
in \kms. The optical depth $\tau$ is derived from the brightness
temperatures of the line $T_{L}$ and the continuum $T_{C}$ by

\begin{equation}
\tau=-\ln\left(1-\frac{|T_{L}|}{T_{C}}\right).
\label{eq:tau}
\end{equation}

\noindent From the different $N_{u}$ of the metastable
$(J=K)$ inversion lines, the rotational temperatures $T_{JJ'}$ can be
derived using

\begin{equation}
\frac{N_{u}(J',J')}{N_{u}(J,J)}=\frac{g_{\rm op}(J')}{g_{\rm op}(J)}\frac{2J'+1}{2J+1}\,\exp\left(\frac{-\Delta E}{T_{JJ'}}\right)
\label{eq:t}
\end{equation}

\noindent \citep[corrected
version of the equation given in][]{hen00}. $\Delta E$ is the energy
difference between the \amm$(J',J')$ and the
\amm$(J,J)$ levels in K [$41.2$\,K between
\amm\ (1,1) and (2,2), and $284.4$\,K between \amm\ (3,3) and (6,6)], 
$g_{\rm op}$ are the statistical weights given as $g_{\rm op}=1$ for
para-ammonia [\amm\ (1,1) and (2,2)] and $g_{\rm
op}=2$ for ortho-ammonia [\amm\ (3,3) and (6,6)]. This equation is
applicable for emission and is also valid for absorption lines when it
is assumed that both transitions have the same $T_{\rm ex}$. Our LVG
analysis (see below) shows that excitation temperatures of \amm\ (1,1)
and (2,2) are indeed very similar, which is not necessarily the case
for \amm\ (3,3) and (6,6).

\placetable{tab:T}

\begin{figure}
\epsscale{1}

\plotone{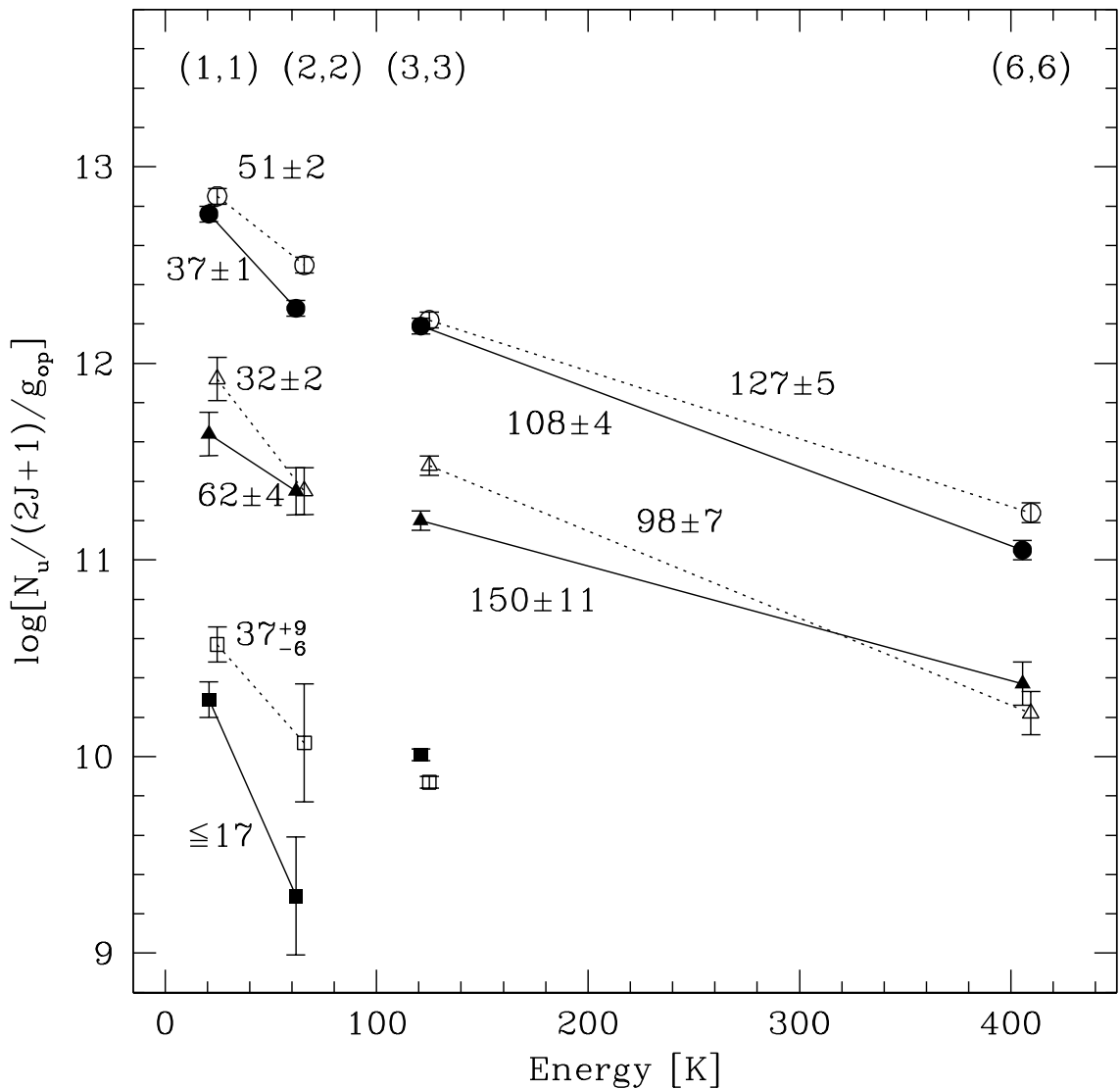}
\caption
{Boltzmann plot of the different ammonia transitions for the NE ({\it
solid lines, filled symbols}) and SW ({\it dotted lines, open
symbols}) molecular complexes. The horizontal axis denotes the
energies of the levels above the ground state. The {\it circles} are
our data, the {\it triangles} are taken from \citet{mau03} and are
{\it shifted down by one dex} on the vertical axis. The {\it squares}
display the results from T02 {\it shifted down by two dex}. The numbers mark
the rotational temperatures in K for the different slopes.
\label{fig:boltzcomp}}

\end{figure}

In Table\,\ref{tab:T} the column densities of the upper levels
($N[J,K]/T_{ex}$ for P4) and the rotational temperatures of the
ammonia spectra at positions P1 to P8 are listed. The column densities
of the para-\amm (1,1) and ortho-\amm (3,3) transitions may differ by
up to a factor of two with a slight trend for $N_{\rm u} (1,1)$ to be
larger. In spite of their higher statistical weights, the (2,2) and
(6,6) levels are less populated than the lower counterparts of the
respective ammonia variant. The rotational temperatures $T_{12}$
(using the para-ammonia [1,1] and [2,2] inversion lines) are in the
range of $35-65$\,K for the emission components and $88$\,K
(statistical 1$\sigma$ error including calibration uncertainties:
$\sim$2\,K) for the absorption component. The rotational temperatures
of ortho-\amm\ ($T_{36}$; using \amm\ [3,3] and [6,6]) are in the
range of $\sim 105-140$\,K (see Table\,\ref{tab:T}). In
Fig.\,\ref{fig:boltzcomp} we compare Boltzmann plots (also referred to
as rotational diagrams) of our analysis with those from the single
dish observations presented by M03 and T02. Since the NE and SW
component are clearly discernible in velocity space, the different
regions can be separated in the single-dish data. Because parts of the
analyses in T02 and M03 are different from ours (e.g., M03 fit a
single rotation temperature using data of all transitions) we
extracted the brightness temperatures from their papers and followed
our methods described above. In the Boltzmann plot, the slopes between
different inversion transitions represent rotational temperatures with
steep slopes corresponding to low rotational temperatures. We find
that our column densities are very similar to those of M03 and exceed
those of T02. T02 did not observe the \amm(6,6) line and could
therefore not derive $T_{36}$. Their $T_{12}$ is lower than ours for
both the SW and the NE components but confirm the trend of higher
temperatures toward the SW. The temperature distribution is reversed
in the analysis of M03. Using HCN and CO data, \citet{pag04}
independently find higher temperatures toward the SW than toward the
NE which agrees with our analysis and that of T02.

\placefigure{fig:boltzcomp}

\subsection{LVG Analysis}

\subsubsection{LVG Models}
\label{sec:lvg}

\begin{figure}
\epsscale{1}
\plotone{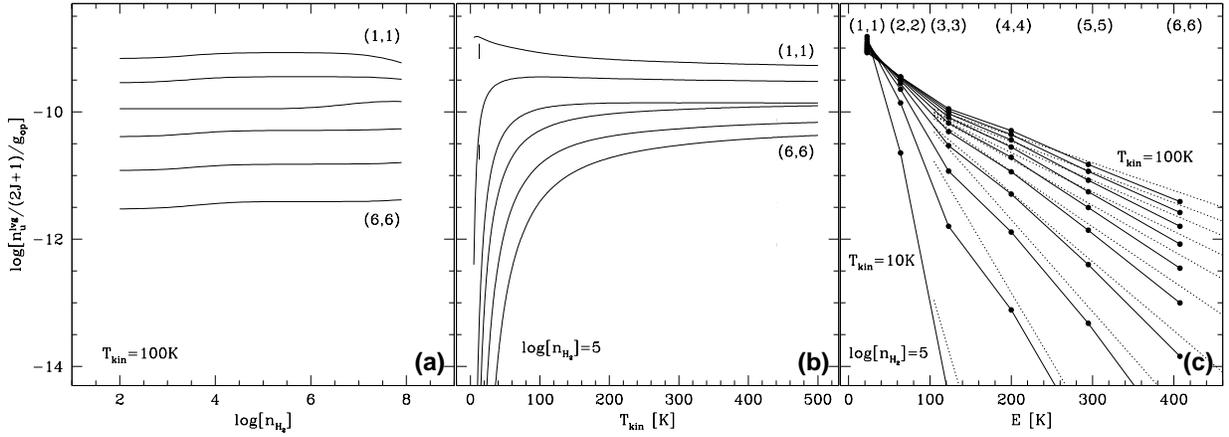}

\caption
{LVG models: Shown are upper level column densities
(corrected for their individual statistical weights as used in
Boltzmann plots) of \amm (1,1), (2,2), (3,3), (4,4), (5,5), and (6,6)
as a function of {\bf (a)} logarithmic \mh\ volume density (kinetic
temperature fixed at $T_{\rm kin}=100$\,K), {\bf (b)} kinetic
temperature (at an \mh\ density of $10^{5}$\,\cden), and {\bf (c)} the
models of (b) as they appear in a Boltzmann plot (kinetic
temperatures: 10--100\,K, spaced by 10\,K). The {\it dotted} lines
mark the expected slopes if $T_{\rm rot}$ would be equal to $T_{\rm
kin}$. The LVG column densities can be converted into real column
densities via eq.\,\ref{eq:lvg}. The models shown are derived for a
fractional \amm\ abundance relative to \mh\ of $10^{-8}$, a velocity
gradient of ${\rm d}v/{\rm d}r=1$\,km\,s$^{-1}$\,pc$^{-1}$, and
$o/p=1$. In panels (a) and (b) the models are displayed for the (1,1)
to (6,6) inversion states from the top to the bottom, respectively.
\label{fig:lvgmodel}}
\end{figure}

As shown by \citet{wal83} and \citet{dan88}, kinetic temperatures and
$T_{12}$ and $T_{36}$ rotational temperatures of NH$_3$ are similar up
to 20 and 60\,K, respectively. Our rotational temperatures, however,
are above those values and therefore the kinetic temperatures exceed
the rotational temperatures considerably. To derive kinetic
temperatures, we applied our LVG models using collisional rate
coefficients and cross sections of \citet{dan88} \citep[see
also][]{sch05}. The LVG models predict column densities for all levels
which depend on the kinetic temperature, the \mh\ volume density, a
velocity gradient, and the temperature of the cosmic microwave
background (CMB). For our calculations we set the CMB temperature to
2.73\,K and the velocity gradient to ${\rm d}v/{\rm
d}r=1$\,km\,s$^{-1}$\,pc$^{-1}$. In the case of optically thin lines,
which is considered here, \amm\ column densities scale linearly with
\amm\ abundances. In Fig.\,\ref{fig:lvgmodel} the column densities in 
the upper (1,1) to (6,6) levels are shown for different models using
an ortho--to--para ($o/p$) ammonia ratio of unity. The densities
($n^{\rm lvg}_{u}$) are displayed in the natural units of the LVG code
which can be converted to measured column densities of the upper
levels $N_{u}$ via

\begin{equation}
N = [n^{\rm lvg}/({\rm d}v/{\rm d}r)]\times3.08\times10^{18}\times\Delta v_{1/2};
\label{eq:lvg}
\end{equation} 

$N$ given in \cden, $n^{\rm lvg}$ in \vden, $\Delta v_{1/2}$ in \kms,
and the velocity gradient ${\rm d}v/{\rm d}r$ in
km\,s$^{-1}$\,pc$^{-1}$.
As shown in Fig.\,\ref{fig:lvgmodel}(a), the computed ammonia column
densities of the metastable lines are virtually independent of the
\mh\ density. Most complex molecules in interstellar space were found
at densities $10^{4}$\,\vden$\lesssim n_{\rm H_{2}}\lesssim
10^{7}$\,\vden\ and we expect that the densities of the environment
from which the ammonia is emitted in NGC\,253 are within that
range. For all further calculations, we therefore fixed $n_{\rm
H_{2}}$ to a value of $10^{5}$\,\vden.

Temperature dependencies are shown in Figs.\,\ref{fig:lvgmodel}(b) and
(c). The fits are most conclusive at low ($\lesssim 50$\,K) kinetic
temperatures where population differences relative to \amm(1,1) column
densities are most sensitive to changes in $T_{\rm kin}$. At higher
temperatures, the ratios between the weighted column densities become
more and more constant and the curves in the Boltzmann plot
(Fig.\,\ref{fig:lvgmodel}[c]) are spaced by smaller intervals. As
expected, the slopes between the data points at a given kinetic
temperature are steeper than those for $T_{\rm rot}=T_{\rm kin}$. In
other words, rotational temperatures underestimate kinetic
temperatures (see above). The most distinctive changes in slope of
$T_{\rm rot}$ are predicted to be at the \amm (3,3) transition and are
most prominent at kinetic temperatures $\gtrsim 50$\,K. Without using
radiative transfer codes, the closest match between $T_{\rm rot}$ and
$T_{\rm kin}$ is therefore provided by all transitions at $T_{\rm
kin}\lesssim 50$\,K, and by the highest states including (3,3) at
$T_{\rm kin}\gtrsim 50$\,K.

\placefigure{fig:lvgmodel}

\subsubsection{Kinetic Temperatures}
\label{sec:T}

\begin{figure}
\epsscale{1}
\plotone{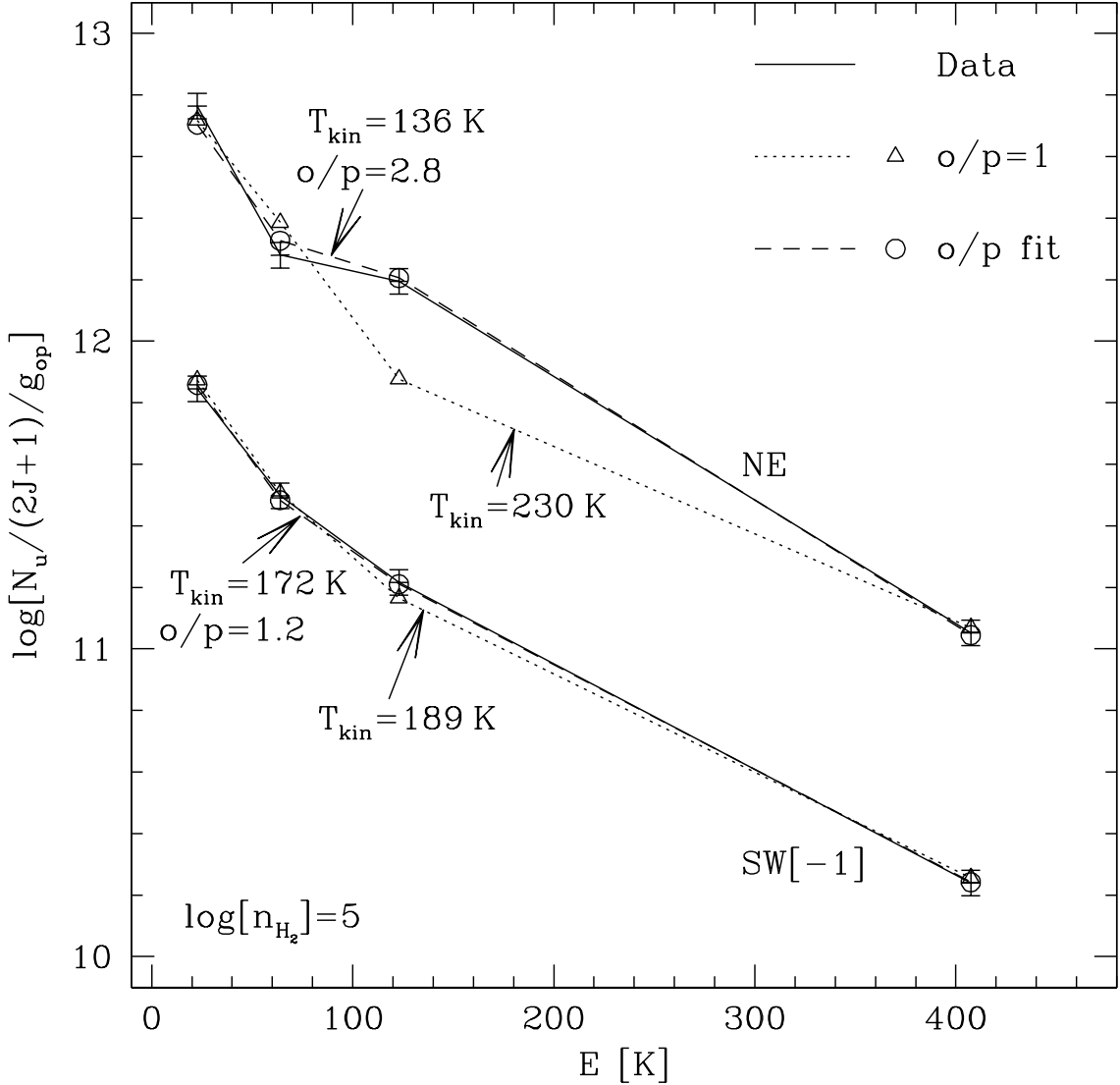}
\caption
{Comparison of one-temperature LVG models with $o/p=1$ and with fitted
$o/p$ ratios for the NE and SW regions. In the Boltzmann plots the
{\it solid lines} connect the data points, whereas the {\it dotted
lines} connect the modelled densities with $o/p=1$ ({\it
triangles}). The {\it dashed lines} connect the results of the LVG
fits with a variable $o/p$ ({\it circles}). All points for the SW
region are {\it shifted down by one dex}. Note that the models with
varying $o/p$ and the data points are very similar, in particular for
the \amm(3,3) measurements. The results of the different fits are
labeled in the plot (see also
Table\,\ref{tab:lvg}). \label{fig:onettwot}}

\end{figure}

We attempted to fit the LVG models to the measured ammonia transitions
at the different positions toward NGC\,253. To do so, we applied one-
and two-temperature models to the data. The second temperatures in the
latter models, however, are usually very similar to the first
temperatures and are therefore not required. One-temperature fits are
shown in Fig.\,\ref{fig:onettwot} for the NE and SW regions. In
general the one-temperature fits with $o/p=1$ were underestimating the
\amm(3,3) column densities but provide reasonable fits to the SW spectra.
Still, the fits are substantially improved when the $o/p$ ratio is
treated as a free parameter. In that case, the data and the models are
virtually indistinguishable. The results of the fits are listed in
Table\,\ref{tab:lvg}. We estimate the errors of the fits including the
free $o/p$ parameter to $\sim 20$\%.

\placefigure{fig:onettwot}
\placetable{tab:lvg}

\begin{figure}
\epsscale{1}
\plotone{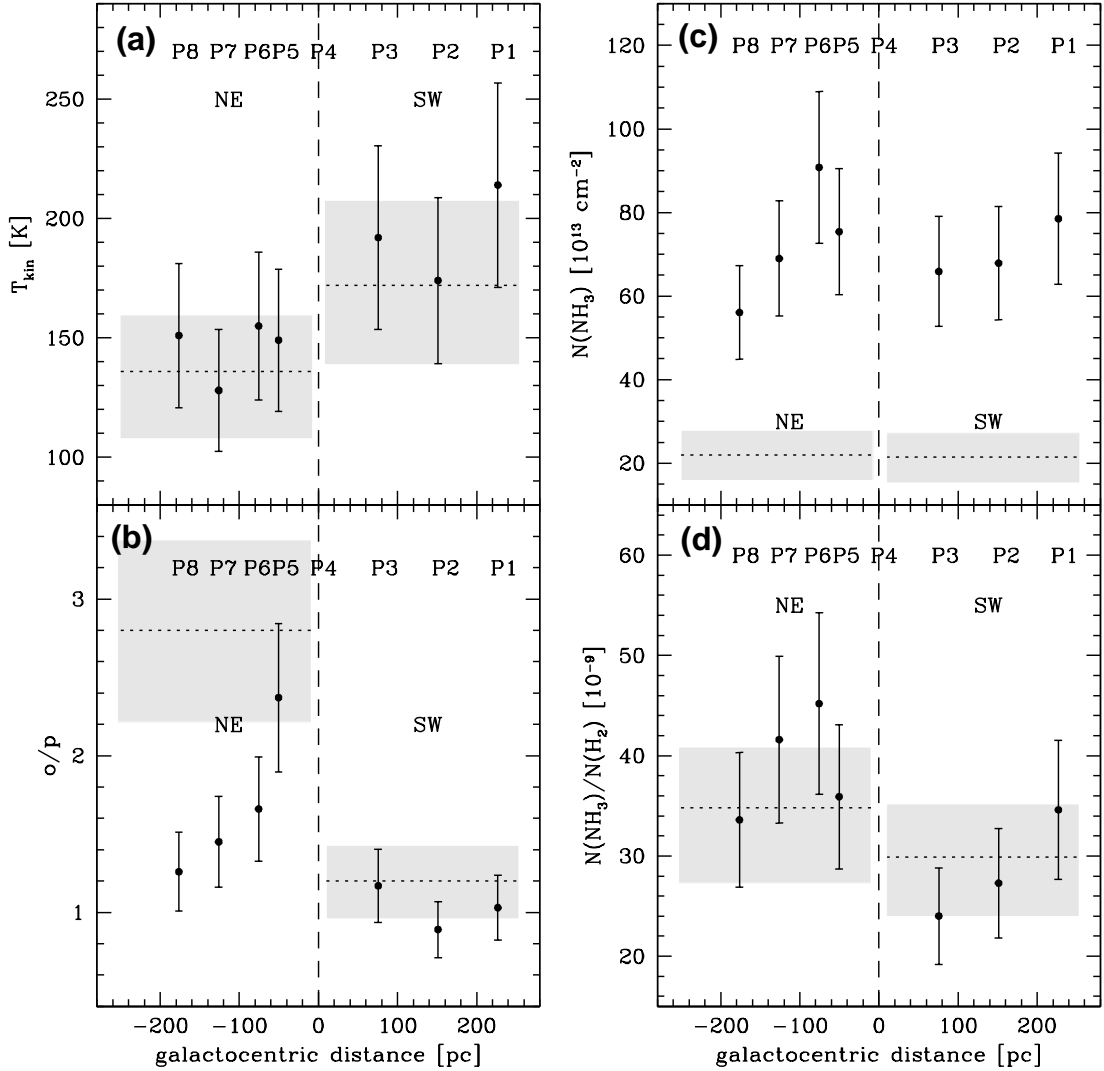}

\caption
{The results of the LVG fits to the data (see Table\,\ref{tab:lvg}) as
a function of galactocentric distance centered on the continuum peak
of NGC\,253 ({\it vertical dashed line}). The ordinates are: {\bf (a)}
the kinetic temperature, {\bf (b)} the $o/p$ ratio, {\bf (c)} the
total ammonia column density, and {\bf (d)} the fractional ammonia
abundance relative to \mh. The {\it dotted horizontal lines} mark the
values derived for the entire NE and SW regions which also incorporate
more diffuse emission. The errors of those values are displayed by the
{\it grey boxes}. \label{fig:lvg}}
\end{figure}

The resulting kinetic temperatures are $\sim 200$\,K in the SW and
$\sim 140$\,K in the NE. Those values are in agreement with the
analysis of \citet{rig02}. They use infrared ISO--SWS data to
determine H$_{2}$ excitation temperatures of the molecular gas. In
addition to two phases with much higher temperatures, they detect
\mh\ at a temperature of $\sim 200$\,K. Based on {\it Spitzer} and 
near--infrared data, \citet{dev04} and \citet{pak04} show that toward
the SW the strengths of the \mh (0--0)\,S(1) and (1--0)\,S(1) lines
are larger than those toward the NE, indicating a larger amount of
warm gas toward the SW. This result corroborates our derived
temperature difference between the NE and SW regions. Temperatures in
the observed range may be explained by shock heating as proposed for
the central Galactic region \citep[e.g.,][]{flo95,hue95,mar01} and
also for NGC\,253 itself and other starburst galaxies
\citep[e.g.,][]{gar00,dev04,pak04}. Other processes, such as 
ion-slip or cosmic ray heating, as well as dynamic heating by the bar
may also have an influence on the temperature of the dense gas. The
kinetic temperature as a function of galactocentric distance are
displayed for the individual positions P1 to P8 and the entire NE and
SW regions in Fig.\,\ref{fig:lvg}(a). Whereas the temperature
variations within the SW and NE regions are not significant, the
temperature difference between the two sides of the nucleus is
notable. Some additional heating may therefore have occurred in the SW
of NGC\,253. For possible mechanisms, see the end of
Sect.\,\ref{sec:n} and Sect.\,\ref{sec:shell}.

\placefigure{fig:lvg}

\subsubsection{The Ortho--to--Para Ammonia Ratio}
\label{sec:op}

The $o/p$ ratio depends on the energy that is transferred to the \amm\
molecule during its formation and equilibration (see
Sect.\,\ref{sec:intro}). At low energies, only the lowest (0,0) state
can be populated which belongs to ortho-ammonia. This results in an
increase of the $o/p$ ratio. Due to the very slow decay of the
metastable inversion states, the $o/p$ ratio is not significantly
altered once the formation and equilibration processes are completed
\citep[see, e.g.,][]{che69,ho83}. As shown in T02, the theoretical
$o/p$ ratio is about unity at formation temperatures above $\sim
30$\,K and $o/p$ rises steeply at lower temperatures. The $o/p$ ratio
therefore offers an archaeological view on the initial conditions of
the molecular gas during its formation.

We derive $o/p$ ratios of $\sim 1$ in the SW region, which would
correspond to an \amm\ formation temperature above $\sim 30$\,K
(cf. fig.\,3 in T02). In the NE region $o/p$ is with $\sim 1.5-2$
significantly larger (Table\,\ref{tab:lvg},
Fig.\,\ref{fig:lvg}b). Such $o/p$ ratios suggest lower ammonia formation
temperatures of $\sim 15-20$\,K toward the NE. This result is in
agreement with the derived kinetic temperatures which are also higher
in the SW region.

Since LVG models are rarely applied when analyzing extragalactic
\amm\ emission, $o/p$ ratios are usually calculated in a different
way making use of deviations from a constant $T_{12}$ rotational
temperature model (a single line in the Boltzmann plot; see, e.g.,
T02; \citealt{tak00}). As shown in Fig.\,\ref{fig:onettwot} this
approach may provide reasonable $o/p$ values when adjusting the
ortho-\amm(3,3) column densities to the extrapolated (para-\amm)
$T_{12}$ temperature (note that the kinetic temperatures of the LVG
models with a fixed $o/p=1$ and a free $o/p$ are similar,
Table\,\ref{tab:lvg}). This approach fails, however, when the
ortho-\amm (6,6) column densities are scaled to match the extrapolated
$T_{12}$. The best method without the application of radiative
transfer models is to adjust para-\amm (4,4) or (5,5) weighted column
densities to $T_{36}$ (see Fig.\,\ref{fig:lvgmodel}[c]).

\subsubsection{Ammonia Column Densities, Abundances, and Masses} 
\label{sec:n}

Total ammonia column densities are calculated by adding up the column
densities derived for the populations of the individual rotational
levels. A rough estimate is given by the sum of the {\it observed}
levels $N_{\rm 1236}({\rm NH_{3}})\equiv2\,\sum_{J=1,2,3,6}
N_{u}(J,J)$ (Table\,\ref{tab:T}). The factor of 2 has been introduced
to accommodate the populations of the lower inversion levels (which
contribute equally to the column densities due to the low energy
difference between upper and lower states of $\sim 1$\,K). A map of
the spatial $N_{\rm 1236}({\rm NH_{3}})$ distribution is shown in
Fig.\,\ref{fig:ntot}. The column densities for the individual clumps
are in the range of $\sim 30-50\times 10^{13}$\,\cden\
(Table\,\ref{tab:T}). Averaged over the entire NE and SW regions,
however, the \amm\ column densities drop to $\sim 10\times
10^{13}$\,\cden\ toward the NE and to $\sim 13\times 10^{13}$\,\cden\
toward the SW, due to the inclusion of large-scale diffuse emission
between the individual complexes A to F. Therefore, both regions
within NGC\,253 exhibit similar averaged \amm\ columns.

Using only the observed values, however, underestimates the true
column densities due to the contributions of the other ammonia
levels. The LVG models predict that collisional excitation of the
non-metastable levels leads to populations about three orders of
magnitudes below those of metastable levels. The column densities of
the non-metastable levels may, however, increase substantially by
infrared pumping \citep[e.g.,][]{mau85} which is not incorporated in
our code. Among the metastable levels, the most prominent are the \amm
(0,0), (4,4) and (5,5) levels. Adding up all the levels involved in
our LVG code (up to the metastable [6,6] levels) the total \amm\
column densities are $\sim 1.5-2.5$ times larger than $N_{\rm 1236}$,
corresponding to column densities of $6-11\times 10^{14}$\,\cden\ for
the individual clumps (Table\,\ref{tab:lvg}). The column densities
averaged over the NE and SW regions are about $2-3$ times lower than
the column densities in the peaks.

Ammonia masses within the NE and SW regions are derived to $\sim 9$
and $\sim 10$\,\msun, respectively, and are therefore comparable. They
add up to a total ammonia mass of $\sim 20$\,\msun\ within the central
2\arcmin\ of NGC\,253 (see Table\,\ref{tab:lvg}).

\amm\ abundances are computed relative to the column densities of \mh. 
The latter are derived using the CO data (see Sect.\,\ref{sec:morph})
with a CO-to-\mh\ conversion factor of $X_{\rm
CO}=5\times10^{19}$\,\cden\,(K\,\kms)$^{-1}$. This value was
determined for ultraluminous far infrared galaxies by
\citet{dow98} and is widely used for starburst galaxies in the literature. 
For NGC\,253 itself, \citet{mau96} suggest a similar value of $X_{\rm
CO}\sim3\times10^{19}$\,\cden\,(K\,\kms)$^{-1}$ based on single-dish
CO data. Note, however, that $X_{\rm CO}$ is not necessarily constant
over the entire starburst. Using multi--transition, interferometric CO
observations toward the starburst galaxy M\,82, \citet{wei01b} showed
that $X_{\rm CO}$ varies in the range of 3 to
10$\times10^{19}$\,\cden\,(K\,\kms)$^{-1}$ along the major axis.

\begin{figure}
\epsscale{0.5}

\plotone{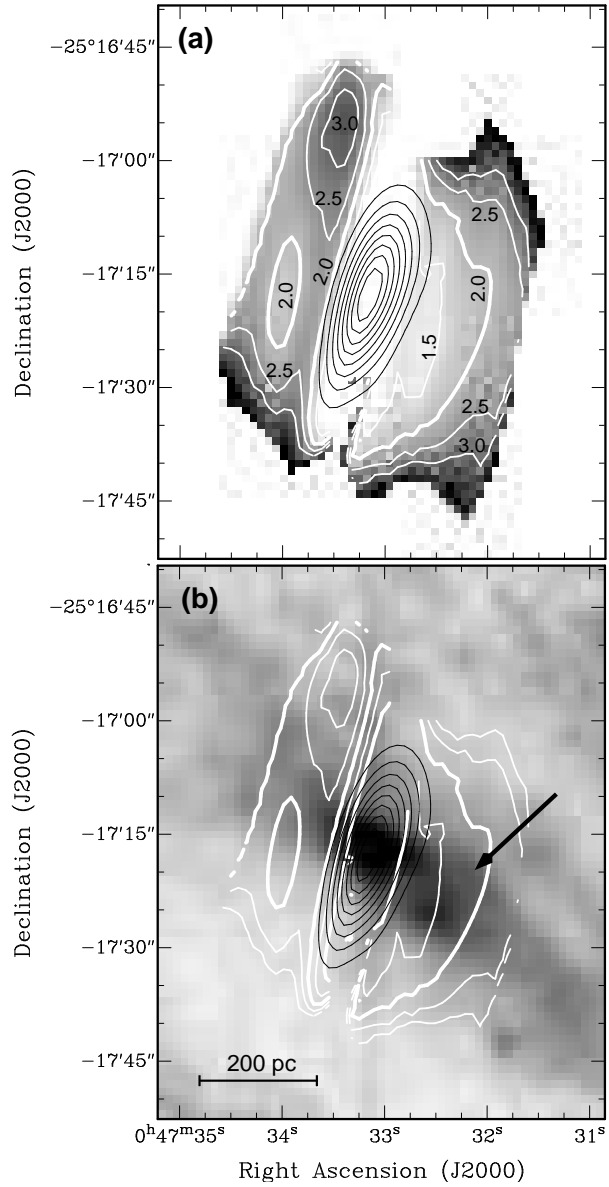}
\caption
{{\bf (a)} Ammonia abundance map of NGC\,253. The {\it white contours}
are the ammonia abundances with respect to \mh\ in units of $10^{-8}$
(contours start at $1.0\times10^{-8}$ and are spaced by
$0.5\times10^{-8}$). The {\it black contours} represent the 1.2\,cm
continuum emission. Note that this map is based on $N_{\rm 1236}({\rm
NH_{3}})$ which underestimates the true abundances by $\sim 1.5-2.5$
(see Sect.\,\ref{sec:n} and Fig.\,\ref{fig:ntot}). The same contours
are overlaid on a 2MASS $J-K$ color map in panel {\bf (b)} where dark
colors are brighter in the K-band. The {\it black arrow} marks the
position of the expanding shell around the X-ray point source shown in
Fig.\,\ref{fig:shell}.
\label{fig:abund}}

\end{figure}

The ammonia abundances derived toward the different positions in
NGC\,253 are listed in Table\,\ref{tab:lvg} and are shown as a
function of galactocentric distance in Fig.\,\ref{fig:lvg}(d). We also
produced a map of the ammonia abundances relative to H$_2$ (based on
$N_{\rm 1236}[{\rm NH_{3}}]$; see the caveats above) which is shown in
Fig.\,\ref{fig:abund}. The fractional ammonia abundances vary in the
range of $\sim 25-45\times 10^{-9}$. The overall abundances are only
slightly larger than those derived for NGC\,253 by M03 ($\sim 20\times
10^{-9}$). Typical ammonia abundances of Galactic molecular clouds are
with $\sim 30\times10^{-9}$ in the same range \citep[e.g.,
][]{irv87,wal93}. Note that toward the Galactic Center, the ammonia
abundances are at least as large as toward Galactic interstellar
clouds \citep[e.g., ][]{flo95,goi04}.

\placefigure{fig:abund}

An \amm\ abundance gradient is visible along the SW complex with
minimal values near the starburst center (Figs.\,\ref{fig:lvg}(d) and
\ref{fig:abund}; note that the abundance between P6 and P5, 
approaching the center from the NE, is also decreasing). The abundance
toward the SW varies from $\sim 2.5\times 10^{-8}$ to $\sim 3.5\times
10^{-8}$ at distances of $\sim 40$\,pc and $\sim 250$\,pc from the
nucleus, respectively. A similar gradient has been found for M\,82 by
\citet{wei01a}, varying from $\sim 2\times10^{-10}$ close to the
starburst center to $\sim 6\times10^{-10}$ at larger radii. Those
values, however, are about two orders of magnitude lower than the
ammonia abundances in NGC\,253. The abundance gradient may be
explained by the destruction of \amm\ by photo-dissociation near the
starburst similar to what has been proposed for M\,82 by
\citet{wei01a}. Whereas the entire ammonia abundance is raised by
liberating \amm\ from grains by shocks (see also Sect.\,\ref{sec:T}),
close to the nucleus of NGC\,253 the molecules may be destroyed by
photons with energies above $\sim 4.1$\,eV, the dissociation energy of
\amm\ \citep{sut83}. {\it Spitzer} data of [\ion{Ne}{2}] and [\ion{Ne}{3}] 
toward NGC\,253 show that most of the ionizing flux is produced close
to the starburst center, probably with a slight offset toward the SW,
and decreases with galactocentric radius \citep[][note that radiation
hardens with increasing radius but, due to the low dissociation
threshold of ammonia, this is only a second order effect to the
destruction of \amm]{dev04}. This can potentially explain the \amm\
abundance gradient. The molecular gas, however, may be shielded
against the UV photons by dust. A 2MASS $J-K$ color map
(Fig.\,\ref{fig:abund}[b]) shows that there is only little
radial variation of the optical and infrared absorption caused by dust
along the molecular condensation toward the SW. Shielding should
therefore be a relatively constant parameter and the ammonia abundance
gradient remains a function of ionizing photons emerging from the
starburst center.

The suggestion that the SW is more affected by photo--dissociation
than the NE may also be corroborated by infrared observations toward
the central region of NGC\,253. With respect to the nucleus
\citep[position (J2000): $\alpha$=00$^{h}$\,47$^{m}$\,33\fs174,
$\delta$=$-$25\degr\,17\arcmin\,17\farcs08;][based on VLA 1.3\,cm
data]{ulv97}, the infrared peak is shifted by 2\arcsec-3\arcsec\
toward the SW \citep{ket93,sam94,boe98}. While this corresponds only
to a fraction of the ATCA beam at 1.2\,cm, it may nevertheless
indicate that the intensity of SF is increased toward the SW of the
nucleus of NGC\,253, which heats the molecular gas as probed by \amm,
and may dissociate a large fraction of the ammonia molecules (see also
Sect.\,\ref{sec:shell}).

\subsection{An Expanding Shell in the Dense Gas}
\label{sec:shell}

\begin{figure}
\epsscale{1}

\plotone{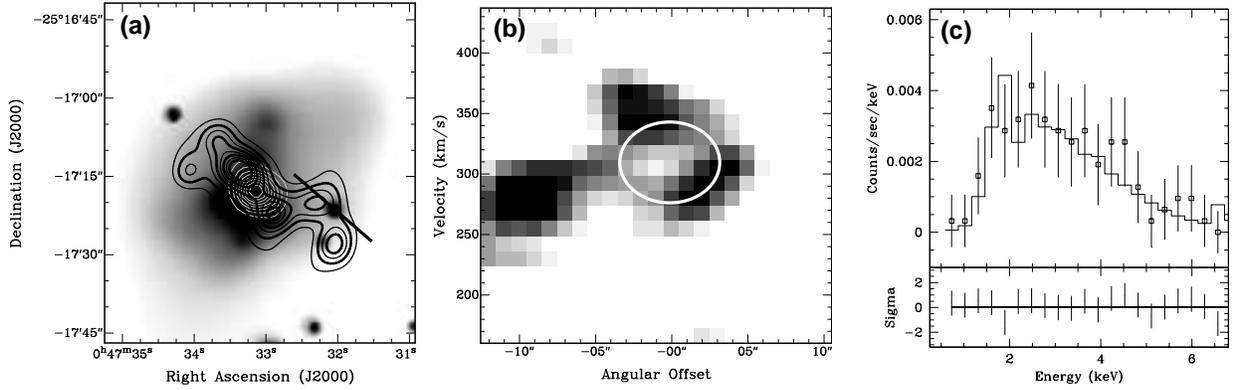}
\caption
{{\bf (a)} The super-resolved \amm (3,3) integrated intensity and
1.2\,cm continuum map (as in Fig.\,\ref{fig:pv}) overlaid as contours
on an adaptively smoothed 0.3--6.0\,keV {\it Chandra} ACIS--S3 image
of NGC\,253. The position velocity (pV) diagram of the \amm (3,3)
data along the line positioned on the X-ray point source toward the
west is displayed in panel {\bf (b)}. In the pV diagram an expanding
shell feature is visible which is marked by the {\it white
circle}. The {\it Chandra} spectrum of the X-ray point source is
displayed in panel {\bf (c)} with a fitted APEC thermal plasma
model. The lower part of this panel shows the errors
([Data-Model]/Error) of the fit.
\label{fig:shell}}
\end{figure}

Toward the northern end of the ammonia emission in the SW region, pV
cuts of the data cubes reveal a feature resembling an expanding shell
that is visible in all inversion lines (see Fig.\,\ref{fig:shell}).
Given the elliptic beam of the radio data, the position of this
feature is somewhat uncertain. However, the largest contrast between
the rim and the center of this shell is found at a location coincident
with an X-ray point source detected in ACIS-S3 images taken with the
{\it Chandra X-ray Observatory} (obs. id 969) at
$\alpha_{(J2000)}=00^h 47^m 32\fs0$, $\delta_{(J2000)}=-25^{\circ}
17\arcmin 21\farcs4$. The X-ray spectrum of this source is shown in
Fig.\,\ref{fig:shell}(c). We fitted an APEC collisional equilibrium
plasma \citep{smi01} with solar metallicity as well as a power law
model to this source
\citep[for the data reduction technique see][]{ott05}. 
Both models provide good fits to the data. The APEC fit results in an
absorbing column density of $N_{\rm
H}=(2.8\pm0.3)\times10^{22}$\,\cden\ (excluding Galactic \hi\
emission), a temperature of $T_{\rm
plasma}=(3.6\pm0.7)\times10^{7}$\,K, a flux of
$F=(1.9\pm0.7)\times10^{13}$\,erg\,cm$^{-2}$\,s$^{-1}$, and an
unabsorbed X-ray luminosity (0.3--6.0\,keV)
$L_{X}=(4.7\pm0.7)\times10^{38}$\,\lum. The power law fit yields
$N_{\rm H}=(2.9\pm0.3)\times10^{22}$\,\cden, a photon index of
$\Gamma=2.3\pm0.1$,
$F=(2.2\pm0.9)\times10^{-13}$\,erg\,cm$^{-2}$\,s$^{-1}$, and
$L_{X}=(10.3\pm2.1)\times10^{38}$\,\lum. In Fig.\,\ref{fig:shell} we
show the APEC fit overlaid on the X-ray spectrum; the power law fit
looks very similar. The X-ray absorbing column density can be compared
to that of molecular hydrogen. From the CO data (resolution used here:
$5\farcs88 \times 2\farcs87$) we derive a luminosity of $\sim
1.1\times10^{3}$\,K\,\kms\ which translates into a proton column
density of $5.5\times 10^{22}$\,\cden, using a CO-to-\mh\ conversion
factor of $5\times10^{19}$\,\cden\,(K\,\kms)$^{-1}$ (see 
Sect.\,\ref{sec:n}). This value is about twice as large as that
derived by X-ray absorption, which suggests that the X-ray point
source is embedded in the molecular material.

As shown in Fig.\,\ref{fig:shell}(b), the shell is centered at a LSR
velocity of $\sim 310$\,\kms\ and expands at $v_{\rm exp}\sim
30$\,\kms. Its radius is $\sim 3''$ which corresponds to $\sim 40$\,pc
at the distance to NGC\,253. The dynamical age of the shell is
therefore $\sim 1.3$\,Myr. Using the shell radius and the proton
column density derived above, we estimate the mass of the shell to
$\sim 8\times10^{6}$\,\msun\ and its kinetic energy to $\sim 7\times
10^{52}$\,erg. This figure is about two (power law model) to four
(APEC) times larger than the energy input of the X-ray source over the
lifetime of the shell. Using the STARBURST99 models \citep{lei99} to
derive the starburst properties \citep[see][ for details]{ott03}, and
assuming that the X-ray luminosity is due to thermalized mechanical
luminosity, the total energy input matches that of a stellar cluster
with a mass of $\sim 10^{5}$\,\msun\ (assuming an instantaneous
starburst, a solar metallicity, and a Salpeter IMF with a lower and
upper mass cutoff of 1 and 100\,\msun, respectively) which must be
highly obscured by the dust visible in the 2MASS $J-K$ map (see
Fig.\,\ref{fig:abund}). The I-band surface brightness at the position
of the cluster candidate (as visible in an uncalibrated F814W
HST/WFPC2 image; proposal ID: 5211) is slightly enhanced (factor of
$\sim 1.2$) as compared to its surroundings. A mass of
$\sim10^{5}$\,\msun\ is typical for globular clusters in the Galaxy
\citep[e.g.,][]{man91} and would also fit the distribution of globular
clusters in the halo of NGC\,253
\citep{bea00}. Note that a young globular cluster candidate of similar
mass has been found in NGC\,253 toward the SW infrared peak
\citep[][]{wat96,ket99}.

\placefigure{fig:shell}

\citet{ulv00} reports a radio continuum source (2\,cm
VLA observations) which is located $\sim 2\arcsec$ toward the east of
the X-ray point source which may be related to the globular cluster
candidate. Within the dynamical timescale of the shell, massive stars
could not have exploded as supernovae. The energy which drives the
shell of dense gas must therefore be provided by massive stellar winds
which shock their ambient medium and thermalize some of it to coronal
temperatures.

For the shell the rotational temperature of the dense gas was
estimated separately using the ammonia observations. Over the area of
the shell, we derive $T_{\rm 12, shell}=(87\pm12)$\,K and $T_{\rm 36,
shell}=(266\pm40)$\,K, which is well a factor of $\sim 2$ above the
mean rotational temperatures averaged over the ammonia emission in the
SW and which converts to extremely high kinetic temperatures. This may
indicate that the energetic input of the super cluster candidate has
heated the surrounding molecular medium. As mentioned in
Sect.\,\ref{sec:T}, \citet{dev04} and \citet{pak04} detect some higher
\mh (0--0)\,S(1) and (1--0)\,S(1) emission toward the SW region as
compared to the NE which coincides with the location of the shell. The
shock produced by the central source and the subsequent expansion of
the shell may therefore have excited the molecular hydrogen and heated
the surroundings. This can potentially explain the higher kinetic
temperatures in the SW region as derived in Sect.\,\ref{sec:T}. Unlike
the starburst near the center of NGC\,253 (see the end of
Sect.\,\ref{sec:n}), the stellar cluster may be too young to
effectively photo-dissociate its environment.

\section{Summary}
\label{sec:summary}

In this paper we present the first interferometric ammonia
observations toward NGC\,253, one of the most nearby starburst
galaxies. Our analysis of the ATCA data reveal the following:

\begin{enumerate}

\item 
Using the unresolved 1.2\,cm continuum emission toward the starburst
nucleus in NGC\,253 (flux: $520\pm52$\,mJy), we derive a SFR of $\sim
2.8$\,M$_{\sun}$\,yr$^{-1}$ which agrees well with earlier FIR
estimates.

\item
Ammonia is detected in NGC\,253 up to a radius of $\sim 250$\,pc
toward the SW and up to $\sim 200$\,pc toward the NE of the nuclear
starburst in NGC\,253. Three individual molecular clumps are
identified within those regions on either side of the nucleus. The
ammonia components are found at velocities of up to $\sim 150$\,\kms\
relative to the systemic velocity of NGC\,253. Very close to the
starburst the north--eastern, innermost \amm\ (3,3) emission peak is
closer to the nucleus than the peaks of the other ammonia lines.

\item 
The line shapes and the morphology of ammonia are similar to those of
other molecular species such as CO, CS, SiO, and
H$^{13}$CO$^{+}$. However, some components traced by those molecules
were not detected in \amm\ which can partly be attributed to the S/N
of our observations.

\item 
Toward the nuclear continuum source we observe \amm (1,1), (2,2), and
(6,6) in absorption but \amm (3,3) in emission. This might indicate
the presence of an ammonia maser.

\item
From our large velocity gradient models we find that for each position
toward NGC\,253 a one-temperature gas with a non-uniform
ortho--to--para ($o/p$) ammonia abundance delivers excellent agreement
with the data, despite of different rotational temperatures $T_{12}$
and $T_{36}$. The kinetic temperatures in the NE region hover around
140\,K. In the SW region they are with about 200\,K significantly
higher.

\item 
Whereas the $o/p$ abundance ratio is $\sim 1$ toward the SW, we derive
$o/p\sim 1.5-2.5$ toward the NE, with the larger values close to the
starburst center. Such $o/p$ ratios correspond to \amm\ formation
temperatures of $\ga$30 and $\sim 15-20$\,K toward the SW and NE,
respectively. Thus, the ammonia formation and kinetic temperatures
show the same trend, being warmer in the SW than in the NE.

\item
Total ammonia column densities are $\sim 3\times 10^{14}$\,\cden\
toward the NE and $\sim 2\times 10^{14}$\,\cden\ toward the SW. For
the individual clumps we derive column densities in the range of $\sim
6-11\times 10^{14}$\,\cden. The total ammonia mass adds up to $\sim
20$\,M$_{\sun}$ with about half of the mass toward either side of the
starforming nucleus. In NGC\,253, ammonia abundances toward the most
prominent molecular complexes are $\sim 2.5-4.5\times 10^{-8}$ with
respect to \mh. Decreasing abundances are measured in the SW complex
toward the starburst center. At the position of the lowest
\amm\ abundance, prominent dust features are visible in near-infrared
2MASS color images. As the starburst in NGC\,253 is slightly shifted
toward the SW, the resulting UV radiation may photo-dissociate the
fragile ammonia molecules predominantly in this region.

\item 
An expanding shell feature is detected within the SW molecular
complex. The shell coincides with a bright X-ray point source. The
absorbing column density of the X-ray source as compared to the total
molecular column reveals that this source is likely located inside the
shell. The shell has a dynamical age of $\sim 1.3$\,Myr and a kinetic
energy of $\sim 7\times10^{52}$\,erg. This corresponds to the energy
input of a stellar cluster with a mass of $\sim
10^{5}$\,M$_{\sun}$. An optical identification of this cluster is very
difficult due to high visual obscuration. At the position of the
shell, the rotational temperature of ammonia is enhanced by a factor
of $\sim 2$ over the mean of the SW region, coincident with a local
maximum of \mh\ excitation temperatures. This supports the scenario
that large amounts of energy are injected into the dense gas.

\end{enumerate}

Our study demonstrates the power of interferometric multi-transition
ammonia observations to constrain in detail the physical conditions of
the various phases of molecular gas in starburst environments. The
application of radiative transfer models to the data allows us to
determine the conditions with better accuracy than previously
possible. Studies of other nearby actively star forming galaxies will
be needed to find out whether NGC\,253 provides a typical environment
and may therefore be used as a prototype for comparisons with more
distant, even more vigorously violently star forming objects.

\acknowledgments
The Australia Telescope Compact Array is part of the Australia
Telescope which is funded by the Commonwealth of Australia for
operation as a National Facility managed by CSIRO. We thank Michael
Dahlem for the provision of the CO data. We are also grateful to
Rainer Mauersberger and the referee for their comments on the
manuscript. C.H. thanks for support provided by ATNF during his time
spent at ATNF and ATCA. This research has made use of the NASA/IPAC
Extragalactic Database (NED), which is maintained by the Jet
Propulsion Laboratory, Caltech, under contract with the National
Aeronautics and Space Administration (NASA), NASA's Astrophysical Data
System Abstract Service (ADS), NASA's SkyView and the NASA/IPAC
Infrared Science Archive, which is operated by the Jet Propulsion
Laboratory, California Institute of Technology, under contract with
the National Aeronautics and Space Administration.

Facilities: \facility{ATCA}, \facility{OVRO}, \facility{Chandra}, \facility{HST}


\clearpage

\clearpage

\clearpage

\clearpage

\clearpage

\clearpage

\clearpage

\clearpage

\clearpage

\clearpage

\clearpage

\clearpage

\clearpage

\clearpage

\begin{deluxetable}{lccc}
\tabletypesize{\small}
\tablecolumns{4}
\tablecaption
{Global line properties of the ATCA ammonia observations for the NE,
SW, and total regions ($v_{\rm LSR,p}$: LSR peak velocity, $\Delta
v_{1/2}$: line FWHM, $\int S\,dv$: integrated flux). The errors of the
velocities are $\sim 20$\,\kms\ ($\sim 40$\,\kms\ for the [6,6] line
due to the lower S/N ratio) and the fluxes are accurate to $\sim 10$\%
(\amm[6,6]: 20\%).
\label{tab:flux}}
\tablewidth{0pt}

\tablehead{Region  &  $v_{\rm LSR,p}$    &  $\Delta v_{1/2}$  &  $\int S\,dv$\\
 & [\kms] & [\kms] & [Jy\,\kms]}

\startdata

\cutinhead{\amm (1,1)} 
NE       &  191    &  94     & 2.00 \\
SW       &  304    &  76     & 2.72 \\             
Total    & \nodata & \nodata & 4.72 \\           
\cutinhead{\amm (2,2)} 
NE       &  178    &  86     & 1.48 \\
SW       &  293    &  66     & 2.70 \\             
Total    & \nodata & \nodata & 4.18 \\           
\cutinhead{\amm (3,3)} 
NE       &  180    &  63     & 3.91 \\
SW       &  304    &  74     & 4.19 \\             
Total    & \nodata & \nodata & 8.10 \\           
\cutinhead{\amm (6,6)} 
NE       &  179    &  60     & 0.60 \\
SW       &  282    &  72     & 1.35 \\             
Total    & \nodata & \nodata & 1.95 \\           
\enddata

\end{deluxetable}

\clearpage

\begin{deluxetable}{llcccccccc}
\tabletypesize{\scriptsize}
\tablecolumns{10}
\tablecaption
{Ammonia and CO line parameters toward positions P1 to P8. The
coordinates are given in the header of the table. The line parameters
are as follows. $T_{\rm mb,p}$: peak main beam brightness temperature,
$v_{\rm LSR,p}$: peak LSR velocity, $\Delta v_{1/2}$: line FWHM, $\int
T_{mb}\, dv$: integrated brightness temperature. Note that for P4 the
parameters are given for the absorption feature of
\amm (1,1), (2,2), (6,6) and for emission of \amm (3,3) and CO(1--0). 
The errors of the velocities are $\sim 20$\,\kms\ ($\sim 40$\,\kms\
for the [6,6] line), the fluxes are accurate to $\sim 10$\%
(\amm[6,6]: 20\%).
\label{tab:clumps}}
\tablewidth{0pt}

\tablehead{Parameter  & Unit&  P1  &  P2  &  P3  &  P4  &  P5  &  P6  &  P7  &  P8}

\startdata
RA~~(J2000) &~~~$00^{h}\,47^{m}$ & 32\fs1 & 32\fs3 & 32\fs9 & 33\fs2 &
33\fs4 & 33\fs5 & 33\fs7 & 34\fs0\\ DEC(J2000) &$-25\degr\,17\arcmin$
& 27\arcsec & 20\arcsec & 21\arcsec & 17\arcsec & 14\arcsec &
14\arcsec & 10\arcsec & 09\arcsec\\
\cutinhead{\amm (1,1)} 
$T_{\rm mb,p}$   & [K]       &  0.16  &  0.15  &  0.13  &  -0.03  &  0.12  &  0.15  &  0.14  &  0.14\\
$v_{\rm LSR,p}$  & [\kms]    &   304  &  330   &  279   &  228    &  165   &  165   &  190   &  203\\
$\Delta v_{1/2}$    & [\kms]    &\phn   76  & \phn 89    &\phn  52    & \phn  32    &\phn  69    & \phn 89    &\phn  81    & \phn 65\\
$\int T_{mb}\, dv$  & [K\,\kms] & 13.38  &  14.37 &  8.43  &  -0.93  &  9.75  &  13.58 &  12.16 &  11.57\\
\cutinhead{\amm (2,2)}
$T_{\rm mb,p}$   & [K]       &  0.15  &  0.12  &  0.14  &  -0.02  &  0.09  &  0.14  &  0.10 &  0.11\\
$v_{\rm LSR,p}$  & [\kms]    &   293  &  331   &  280   &  230    &   179  &  179   &  179  &  205\\
$\Delta v_{1/2}$    & [\kms]    &  \phn67  & \phn 91  & \phn 50  & \phn 52  & \phn 64  & \phn 68  & \phn 63  & \phn 63\\
$\int T_{mb}\, dv$  & [K\,\kms] &  12.23  &  10.98  &  10.03  &  -1.19  &  6.60  &  11.31  &  8.35  &  7.95\\
\cutinhead{\amm (3,3)}
$T_{\rm mb,p}$   & [K]       &  0.21  &  0.15  &  0.28  &  0.15  &  0.31  &  0.28  &  0.20  &  0.17\\
$v_{\rm LSR,p}$  & [\kms]    &  305   &  330   &  280   &   167  &  167   &  167   &  192  &  205\\
$\Delta v_{1/2}$    & [\kms]    &  \phn63  & \phn 89  &\phn  63  &  113  & \phn 67  & \phn 62  & \phn 70  & \phn 51\\
$\int T_{mb}\, dv$  & [K\,\kms] &  16.07  &  13.36  &  17.87  &  20.11  &  22.18  &  20.34  &  16.00  &  10.97\\
\cutinhead{\amm (6,6)}
$T_{\rm mb,p}$   & [K]       &  0.06  &  0.04  &  0.06  &  -0.03  &  0.06  &  0.06  &  0.03  &  0.04\\
$v_{\rm LSR,p}$  & [\kms]    & 290  &  338  &  290  &  230  &  170  &  170  &  158  &  206\\
$\Delta v_{1/2}$    & [\kms]    & \phn 53  & \phn 78  & \phn 41  & \phn 48  & \phn 54  &\phn  66  & \phn 94  &\phn  36\\
$\int T_{mb}\, dv$  & [K\,\kms] &  4.59  &  3.17  &  3.37  &  -1.47  &  3.54  &  4.5  &  3.12  &  1.62\\
\cutinhead{CO(1--0)}
$T_{\rm mb,p}$   & [K]       &  4.15  &  3.90  &  4.00  &  2.62  &  2.95  &  2.67  &  2.32  &  2.47\\
$v_{\rm LSR,p}$  & [\kms]    &  318  &  318  &  268  &  167  &  154  &  167  &  205  &  205\\
$\Delta v_{1/2}$    & [\kms]    &  \phn94  &  114  &  \phn89  &  226  &  138  &  137  &  151  &  139\\
$\int T_{mb}\, dv$  & [K\,\kms] &  454.6  &  497.1  &  549.3  &  465.9  &  420.6  &  401.8  &  331.7  &  334.2\\
\enddata

\end{deluxetable}

\clearpage

\begin{deluxetable}{llcccccccccc}
\tabletypesize{\footnotesize}
\tablecolumns{12}
\tablecaption{Derived quantities of the ammonia inversion lines 
at positions P1 to P8 and the entire NE and SW regions. $N_{\rm u}$
are the column densities of the upper levels, $N_{\rm 1236}({\rm
NH_{3}})$ are the combined measured column densities (assuming that
the lower levels are as populated as the upper levels, but ignoring
all contributions from other lines), $N_{\rm 1236}({\rm
NH_{3}})/N({\rm H_2})$ are the abundances based on the $N_{\rm
1236}({\rm NH_{3}})$ column densities. $T_{12}$ and $T_{36}$ are the
rotational temperatures using the para-ammonia (1,1) and (2,2) and the
ortho-ammonia (3,3) and (6,6) inversion lines, respectively. Whereas
the absolute uncertainties of the column densities are $\sim 10$\%,
(\amm[6,6]: 20\%) the relative uncertainties are $\sim 5$\%. $T_{12}$
and $T_{36}$ exhibit a statistical 1$\sigma$ error of 2 and
5\,K, respectively.\label{tab:T}}

\tablewidth{0pt}

\tablehead{Parameter  & Unit&  P1  &  P2  &  P3  &  P4\tablenotemark{a}  &  P5  &  P6  &  P7  &  P8 & NE & SW}

\startdata
$N_{\rm u}(1,1)$  &  [$\times10^{13}$\,\cden]  & 8.78 & 9.42 & 5.53 &9.9\tablenotemark{a} & 6.39 & 8.91 & 7.98 & 7.59 & 1.74 & 2.10\\
$N_{\rm u}(2,2)$  &  [$\times10^{13}$\,\cden]  & 6.01 & 5.39 & 4.93 & 15.4\tablenotemark{a} & 3.24 & 5.56 & 4.1 & 3.91 & 0.95 & 1.57\\
$N_{\rm u}(3,3)$  &  [$\times10^{13}$\,\cden]  & 6.97 & 5.80 & 7.76 & 8.73 & 9.63 & 8.83 & 6.94 & 4.76 & 2.19 & 2.30 \\
$N_{\rm u}(6,6)$  &  [$\times10^{13}$\,\cden]  & 1.66 & 1.15 & 1.22 & 14.2\tablenotemark{a} & 1.28 & 1.63 & 1.13 & 0.59 & 0.29 & 0.45 \\
$N_{\rm 1236}({\rm NH_{3}})$  &  [$\times10^{13}$\,\cden]  & 46.8 & 43.5 & 38.9 & \nodata & 41.1 & 49.8 & 40.3 & 33.7 & 10.4 & 12.8 \\
$N(\rm H_2)$ &  [$\times10^{21}$\,\cden]   & 22.7 & 24.9 & 27.5 & 23.3 & 21.0 & 20.1 & 16.6 & 16.7 & 6.3 & 7.2 \\
$N_{\rm 1236}({\rm NH_{3}})/N({\rm H_2})$  &  [$\times10^{-9}$]   & 20.6 & 17.5 & 14.2 & \nodata & 19.5 & 24.8 & 24.3 & 20.2 & 16.5 & 17.8\\

$T_{\rm 12}$ & [K] & \phn46 & \phn39 & \phn66 & \phn88\tablenotemark{a}& \phn35 & \phn42 & \phn35 & \phn35 & \phn 37 & \phn 51 \\
$T_{\rm 36}$ & [K] & 138 & 127 &115 & \nodata & 108 & 123 & 117 & 105 & 108 & 127 \\

\enddata 

\tablenotetext{a}
{The column densities listed for the \amm\ (1,1), (2,2) and (6,6)
absorption components are actually the ratio of the column density to
the excitation temperature $N/T_{ex}$ in units of
$10^{13}$\,\cden\,K$^{-1}$ (see Sect.\,\ref{sec:Trot}).}
\end{deluxetable}

\clearpage

\begin{deluxetable}{llccccccccc}
\tabletypesize{\footnotesize}
\tablecolumns{11}
\tablecaption{
Results of the LVG model fits (excluding P4). Listed are the best
parameters with a variable $o/p$ ratio: the kinetic temperature
$T_{\rm kin}$, the ortho--to--para-ammonia abundance ratio ($o/p$),
the total ammonia column densities $N({\rm NH_{3}})$, The ammonia
abundances $N({\rm NH_{3}})/N(H_{\rm 2})$, and ammonia masses $M({\rm
NH_{3}})$. In addition we show the kinetic temperatures which are the
results of fits with an $o/p$ ratio of unity [$T_{\rm
kin}(o/p=1)$]. The errors are estimated to be $\sim 20$\%.
\label{tab:lvg}}
\tablehead{Parameter  & Unit&  P1  &  P2  &  P3  &   P5  &  P6  &  P7  &  P8 & NE & SW}
\tablewidth{0pt}
\startdata
$T_{\rm kin}$ & [K] & 214 & 174 & 192 &  149 & 155 & 128 & 151 &  136 & 172\\
$o/p$         &     & 1.0 & 0.89 & 1.2  & 2.4 & 1.7 & 1.5 & 1.3& 2.8 & 1.2\\
$N({\rm NH_{3}})$   & [$\times 10^{13}$\,cm$^{-2}$] &   79.5 & 64.1 & 71.1 & 106 & 113 & 81.7 & 62.5 &  32.4 & 23.5\\
$N({\rm NH_{3}})/N(H_{\rm 2})$ & [$\times 10^{-9}$] &34.6 & 27.3 & 24.0 & 35.9 & 45.2 & 41.6 &  33.6 & 34.8 & 29.9\\
$M({\rm NH_{3}})$ &[M$_{\sun}$]&  \nodata & \nodata  & \nodata & \nodata  & \nodata  & \nodata & \nodata &9.0 & 9.8 \\ 
\tableline
$T_{\rm kin}(o/p=1)$ & [K] & 218 & 168 & 209 & 236 & 196 & 149 & 166 &  230 & 189\\

\enddata 

\end{deluxetable}

\end{document}